\documentclass{article}
\usepackage{a4wide}

\usepackage[colorlinks=true, linkcolor=blue]{hyperref}
\usepackage{glossaries}
\usepackage{natbib}

\usepackage{graphicx}
\usepackage{amsmath}
\usepackage{amssymb}

\usepackage{xcolor}
\usepackage[normalem]{ulem}
\usepackage{mathtools}
\usepackage{placeins}

\newglossaryentry{atlas}{
	name={atlas},
	description={a collection of charts covering the manifold \cite[p. 12]{Lee2013}}
}

\newglossaryentry{chart}{
	name={chart},
	description={a bicontinuous bijection from a subset of a manifold to an open subset of $\mathbb{R}^N$ \citep[p. 12]{Wald1984} \citep[p. 4]{Lee2013}}
}

\newglossaryentry{congruence}{
    name={congruence},
    description={a family of curves, defined on a subset $\mathcal{U}$ of the manifold, such that each of the events of $\mathcal{U}$ belongs to one and only curve}
}

\newglossaryentry{coordinate_system}{
	name={coordinate system},
	description={a chart \citep[p. 12]{Wald1984}}
}

\newglossaryentry{instantaneous_observer}{
	name={instantaneous observer},
	description={a pair $(\mathcal{P}, u^\alpha_{\mathcal{P}})$, where $\mathcal{P}$ is an event of the manifold $\mathcal{M}$ and $u^\alpha_{\mathcal{P}}$ is a timelike future-directed normalized tangent vector (four-velocity) in $T_\mathcal{P}\mathcal{M}$ \citep[p. 43]{Sachs1977}}
}

\newglossaryentry{isometry}{
        name={isometry},
        description={a transformation of a manifold with metric which leaves the metric tensor invariant}
}

\newglossaryentry{Killing_vector_field}{
	name={Killing vector field},
	description={an infinitesimal symmetry of the metric tensor (isometry), i.e., a solution $K^\alpha(x^\beta)$ 
		of Killing's equation: $K_{(\alpha;\beta)}=0$ \citep[p. 441]{Wald1984}}
}

\newglossaryentry{manifold}{
	name={manifold},
	description={a topological space locally equivalent (diffeomorphic) to some Euclidean $N$-dimensional space \citep[p. 12]{Wald1984} \citep[p. 2, 13]{Lee2013}}
}

\newglossaryentry{observer}{
	name={observer},
	description={a timelike future-directed normalized curve \citep[p. 41]{Sachs1977}}
}

\newglossaryentry{reference_frame}{
	name={reference frame},
	description={a vector field whose integral curves are observers \citep[p. 52]{Sachs1977}}
}

\newglossaryentry{spacetime}{
	name={spacetime},
	description={a pair $(\mathcal{M}, g_{\alpha\beta})$\,, where $\mathcal{M}$ is a differentiable manifold and $g_{\alpha\beta}$ is a Lorentzian metric tensor field defined on $\mathcal{M}$ \citep[p. 27]{Sachs1977}}
}

\newglossaryentry{static_spacetime}{
	name={static spacetime},
	description={a spacetime with a timelike Killing vector field, which is also hypersurface-orthogonal}
}

\newglossaryentry{tetrad}{
	name={tetrad},
	description={an (orthonormal) basis $e^\alpha_{(A)}$ for the tangent space at a given event}
}

\makeglossaries

\begin{document}
 
\title{Twins in relativistic spacetimes: dispelling some misconceptions}
\author{T. A. A. da Câmara\\\emph{IF-UFRJ} \and P. S. Lemos\\\emph{IF-UFRJ} \and M.\ O.\ Calv\~ao\\\emph{IF-UFRJ} \and F.\ D.\ Sasse\\\emph{CCT-UDESC}}

\maketitle

\begin{abstract}
	In contrast to Newtonian physics, there is no absolute time in relativistic (Lorentzian) spacetimes. This immediately implies that two twins may, in general, age at different rates. For this to happen, there must be, of course, some asymmetry between their worldlines, along which the elapsed proper times are evaluated; such asymmetry might not be, however, so intuitively apparent. Our primary objective is to present, in a concise and didactic manner, some lesser-known results and derive novel ones from a modern, geometrical, and covariant standpoint, which aims to clarify the issue and dispel related misconceptions. First, we recall that: (i) the original ``twin paradox'' may be perfectly dealt with in special relativity (physics in a flat spacetime) and does not necessarily involve an accelerated twin. Then we explore the issue of differential aging in general relativity (physics in a curved background), in the prototypical case of the vacuum Schwarzschild spacetime, by considering several pairs of twins. In this context, we show that: (ii) it is not true that a twin which gets closer to the Schwarzschild horizon, by being subject to a stronger gravitational field, where time sort of slows down, should always get younger than a twin that stays further away, in a region of weaker gravitational field, and (iii) it is also false that an accelerated twin always returns younger than a geodesic one. Finally, we argue that (iv) in a generic spacetime, there is no universal correlation between the phenomena of differential aging and the Doppler effect. Two particularly pedagogical resources provided are a glossary of relevant terms and supplementary Python notebooks in a GitHub repository.
\end{abstract}

\section{Introduction}
\label{sec:intro}

In the highly acclaimed \emph{Interstellar} blockbuster movie from 2014, characters Cooper and Amelia visit the ocean world planet Miller, close to the black hole Gargantua, for a few hours, while their fellow Romilly stays in the spaceship \emph{Endurance}, in orbit around the black hole. When they meet again, Romilly has aged about 23 years!\cite{Thorne2014} This is usually explained by stating that, near massive bodies, time slows down. In another classic movie, \emph{Planet of the Apes}, from 1968, a crew of astronauts returns to an Earth 2 millennia older, after a space voyage that lasted for them only about a year! This is usually attributed to the fact that accelerated bodies grow younger as compared to inertial or geodesic ones. In this article, we will show, \emph{inter alia}, that those types of explanation are not universally valid.

The latter issue, related to the \emph{Planet of the Apes} movie, is frequently referred to in the literature as the so-called ``twin paradox''. We would like, from the outset, to spell out a misunderstanding as regards the use of the term ``paradox'': of course, there is no legitimate logical paradox at all, but only a seemingly contradictory result, as viewed from a Newtonian or special relativistic framework (cf. Secs. \ref{sec:differential_aging} and \ref{sec:SR}). 

In this article, we study the relative aging of several relevant pairs of twins, briefly in special relativity (SR) and more systematically in the particular context of vacuum Schwarzschild spacetime of general relativity (GR), and thereby we provide counterexamples to two recurring impressions that students might infer from the popularization literature or even some textbooks: (1) that a geodesic twin always gets older than an accelerated one (e.g. \cite[p. 34]{ONeill1995}, \cite[p. 21]{Stephani2004}, \cite[p. 3-4]{Taylor2000}), and (2) that a twin which travels closer to a matter source necessarily gets younger than one which stays further away (e.g. \cite[p. 253]{Gleiser2005}, \cite[p. 55, 118]{Rezzolla2023}, \cite[p. 7, 8 and 33]{Rovelli2018}.

The structure of the article is as follows. In Sec. \ref{sec:differential_aging},  in a relativistic context, we present some fundamental concepts related to the loose (and otherwise ambiguous) notion of an observer, show their relevance for the correct specification of any idea of time interval, and present the concrete hypotheses underlying the differential aging issue and its resolution in an arbitrary spacetime. In Sec. \ref{sec:SR}, we very briefly recall a flat spacetime where the asymmetry that explains the differential aging between two smooth observers may not be ascribed to differences in (proper) acceleration. In Sec. \ref{sec:GR}, we describe the four kinds of (intersecting) twins we will consider in vacuum Schwarzschild spacetime: an accelerated static Killing observer and three (not necessarily smooth) geodesic observers. In Sec. \ref{sec:comparison_proper_times}, we derive the ratios of proper times and display the corresponding plots. In Sec. \ref{sec:Doppler_effect}, we argue that, given two arbitrary twins in a generic (curved) spacetime, there is no clear-cut relation between their differential aging and the corresponding Doppler shift between them. Finally, in Sec. \ref{sec:discussion}, we conclude with a discussion of the main results and some future perspectives.

Some of our conventions are: (i) the signature for the metric is ``mostly positive'', i.e., $(-1,1,1,1)$\,; (ii) we will systematically set Newton's gravitational constant $G_N$ and the vacuum speed of light equal to one: $G_N \equiv c\equiv 1$\,; (iii) Greek indices run from 0 to 3, and Latin ones from 1 to 3.

\section{Lorentzian spacetimes and differential aging in general}
\label{sec:differential_aging}

Newtonian physics is developed such that there is, in a given background \gls{manifold}\footnote{All (blue) italicized expressions are defined in a glossary at the end of the article.}, a notion of absolute time, which may, naively speaking, be identified with a privileged scalar field in such a manifold\footnote{In fact, the geometrical structure of the classical (nonrelativistic) spacetimes is, somewhat surprisingly, more complex than the Lorentzian relativistic ones; cf. e.g. \cite{Ehlers1973, Earman1989, Penrose2007}.}. Thus, given two events, $\mathcal{P}$ and $\mathcal{Q}$\,, there is an absolute time interval assigned to them, $\Delta T(\mathcal{P}, \mathcal{Q})\coloneqq T(\mathcal{Q}) - T(\mathcal{P})$\,, irrespective of the observers ($\mathcal{O}_1\,,\mathcal{O}_2\,,\ldots$) which are present there. This expresses the absence, or rather the meaninglessness, of any differential aging (or ``twin paradox'') in such a setting.

With the Newtonian case briefly addressed, we now turn to a Lorentzian  \gls{spacetime}. Following the lead of \cite{Sachs1977}, we urge the reader to distinguish among the concepts of \gls{instantaneous_observer}, \gls{observer}, \gls{reference_frame}, (not to be confused with either a \gls{coordinate_system} or a \gls{tetrad} (field)), which are crucial for a definition of any kind of time interval, and therefore for a perfect understanding of the issue of differential aging (or, as an abuse of language, ``twin paradox'').

Let there be given two infinitesimally close events $x^\alpha$ and $x^\alpha + dx^\alpha$\,. Whenever we want to refer to an infinitesimal time interval measurement between those events, be it in the context of special relativity (locally flat, Minkowski \gls{spacetime}) or general relativity (curved \gls{spacetime} proper), there must be, at least implicitly, an associated \gls{instantaneous_observer} $(\mathcal{P}\,, u^\alpha)$\,, such that the mentioned time interval refers to the proper time assigned by this very instantaneous observer to the infinitesimal displacement $dx^\alpha$, according to:
\begin{equation}
	dT(dx^\alpha, u^\alpha) \coloneqq -dx^\alpha u_\alpha\,. \label{infinitesimal_time_interval}
\end{equation}
This is to be contrasted, for instance, with the curvature tensors (Riemann, Ricci, Weyl, and Einstein), at a given event, which depend only on the metric and its derivatives, but not on any \gls{instantaneous_observer}. 
Likewise, if we want to refer to the time interval between two events $\mathcal{S}$ and $\mathcal{A}$\,, along a given \gls{observer} $\mathcal{O}$\,, parametrized as $\gamma^\alpha(\tau)$\,, we mean, of course:
\begin{equation}
	\Delta\tau[\mathcal{O}_{\mathcal{S}\to{\mathcal{A}}}] \coloneqq \int_{\mathcal{O}_{\mathcal{S}\to \mathcal{A}}} \sqrt{-g_{\alpha\beta}\dot{\gamma}^\alpha(\tau)\dot{\gamma}^\beta(\tau)}\,d\tau\,. \label{time_interval_along_observer}
\end{equation}

\begin{figure}
	\centering
	\includegraphics[scale=0.4]{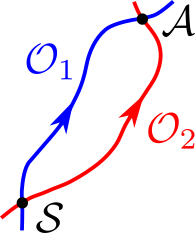}
	\caption{Two relevant observers, ${\mathcal O}_1$ and ${\mathcal O}_2$, meeting at events ${\mathcal S}$ (start) and ${\mathcal A}$ (arrival), for the setting of the differential aging issue in an arbitrary spacetime $({\mathcal M}, g_{\alpha\beta})$.} 	\label{fig:observers_for_twin_paradox}
\end{figure}

The concrete setting for the differential aging issue, in any given \gls{spacetime}, demands, for its very formulation, the consideration of two arbitrary \glspl{observer}, $\mathcal{O}_1$ and $\mathcal{O}_2$\,, geodesic or accelerated, which meet, at least, at two events, $\mathcal{S}$ and $\mathcal{A}$ (cf. Fig. \ref{fig:observers_for_twin_paradox}), so that they can compare their corresponding elapsed proper times between those events, $\Delta\tau_{\mathcal{O}_{1,\mathcal{S}\to\mathcal{A}}}$ and $\Delta\tau_{\mathcal{O}_{2,\mathcal{S}\to\mathcal{A}}}$\,, and thus ascertain whether they are different, i.e., whether there is any differential aging; if that is the case, it would be expedient to somehow account for it. 

The generic explanation, in any \gls{spacetime}, for such a Newtonian unexpected result for the twins, rests simply on three premises:
\begin{enumerate}
	\item the validity of the clock hypothesis, as most clearly exposed by Rindler \cite{Rindler2006a}, which states that the physically relevant time is defined by the proper time of ideal (point) clocks independently of their geodesic or accelerated motion and of their location. This amounts to disregarding any explicit, direct modification of the inner workings of an ideal clock, such as an atomic one, upon the curvature tensors of the underlying neighborhood. Due to the typical dominance of the electric force over the gravitational tidal forces, this is practically always valid {\cite{Parker1982}};
	\item the non-exact character of the infinitesimal interval $dT$ in Eq. \eqref{infinitesimal_time_interval} or, equivalently, the dependence of the elapsed proper time integral $\Delta\tau$ in Eq. \eqref{time_interval_along_observer} on the \gls{observer} connecting two given, fixed events. Alternatively, in contrast to the Newtonian case, there is no absolute time assigned to events (apart from a universal offset);
	\item the ultimate existence of an asymmetry (metrical or topological) between the twins, which implies distinct values for Eq. \eqref{time_interval_along_observer}.
\end{enumerate}

\section{Special relativity: flat spacetimes}
\label{sec:SR}
\begin{figure}[ht]
	\centering
	%\hspace*{-10mm}
	\includegraphics[scale=0.22]{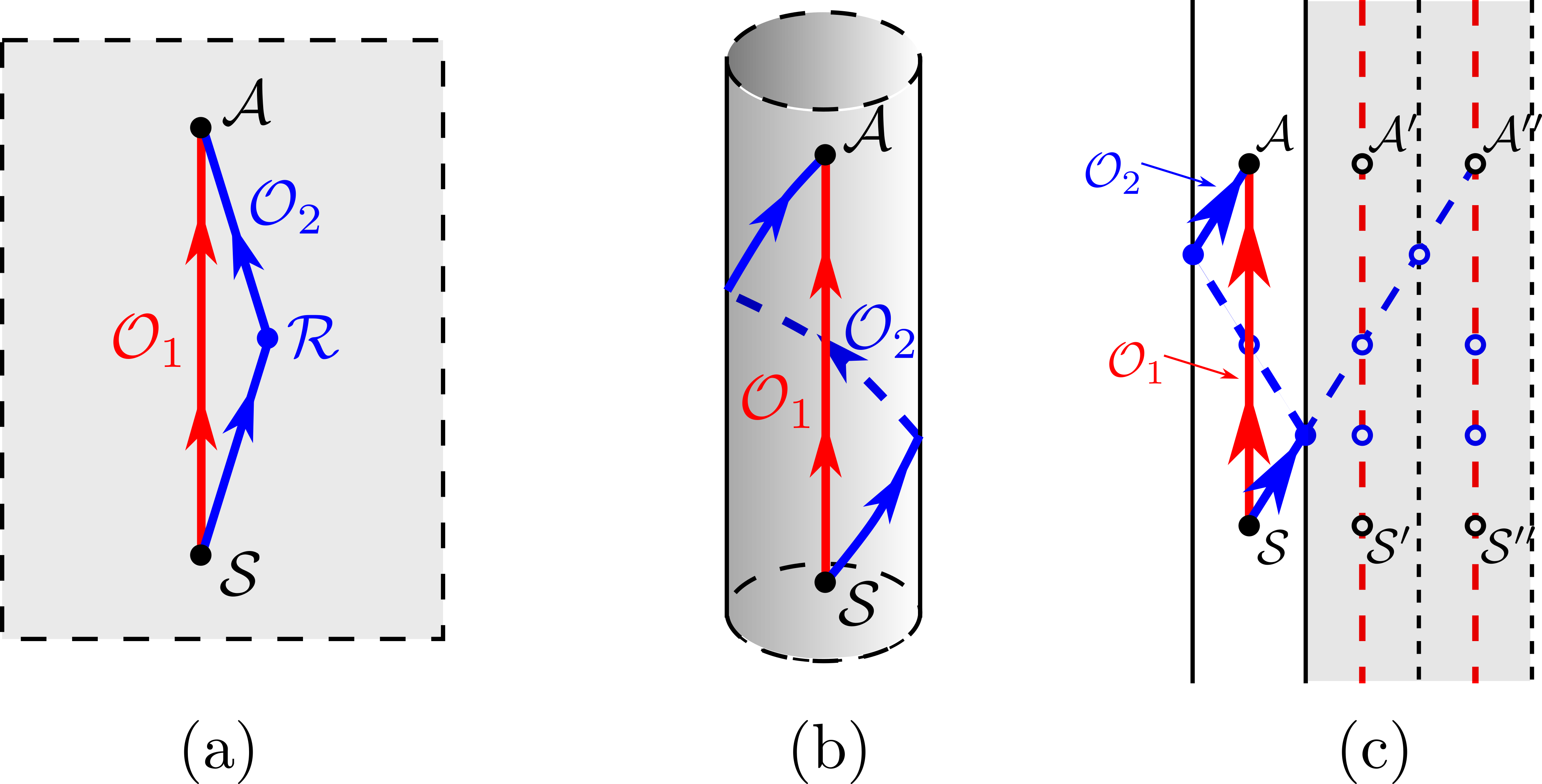}
	\caption{Flat spacetimes: (a) standard, simply connected, Minkowski spacetime. (b) cylindrical Minkowski spacetime (c) covering flat spacetime corresponding to (b).} 	\label{fig:flat_spacetimes}
\end{figure}

By special relativity, we mean physics in a \gls{spacetime} with vanishing Riemann curvature tensor, i.e., in a flat spacetime (usually called a Minkowski spacetime). The underlying manifold may have the trivial topology of $\mathbb{R}^4$ or an ``exotic'' one. If one chooses the trivial one, as depicted for two dimensions in panel (a) of Fig. \ref{fig:flat_spacetimes}, it is obvious that, given two events, $\mathcal{S}$ and $\mathcal{A}$ in its future, then there is a unique (smooth) geodesic \gls{observer} connecting such events, $\mathcal{O}_1$ in the panel, for which the proper time interval Eq. \eqref{time_interval_along_observer} is maximal: any other \gls{observer}, such as a broken (non-smooth) geodesic ($\mathcal{O}_2$ in the panel) or a (smoothly) accelerated one, ages more than $\mathcal{O}_1$\,.  Thus, in this case, the asymmetry that explains the differential aging is due to an acceleration (impulsive or not). Sometimes, people try to extrapolate this to the effect that: (i) an acceleration is always necessary for the asymmetry and (ii) an \gls{observer} which is a smooth geodesic always ages less than any other one (be it a broken geodesic or a smoothly accelerated one). This can be readily refuted by considering the cylindrical Minkowski \gls{spacetime} in panels (b) and (c) of Fig. \ref{fig:flat_spacetimes} [cf. \cite{Brans1973, Dray1990, Barrow2001, Uzan2002}]. Indeed, the observer $\mathcal{O}_2$ in such panels is also a smooth geodesic and, of course, ages less than observer $\mathcal{O}_1$.

\section{General relativity: vacuum Schwarzschild spacetime}
\label{sec:GR}

\begin{figure}[ht]
	\centering
	\includegraphics[scale=0.25]{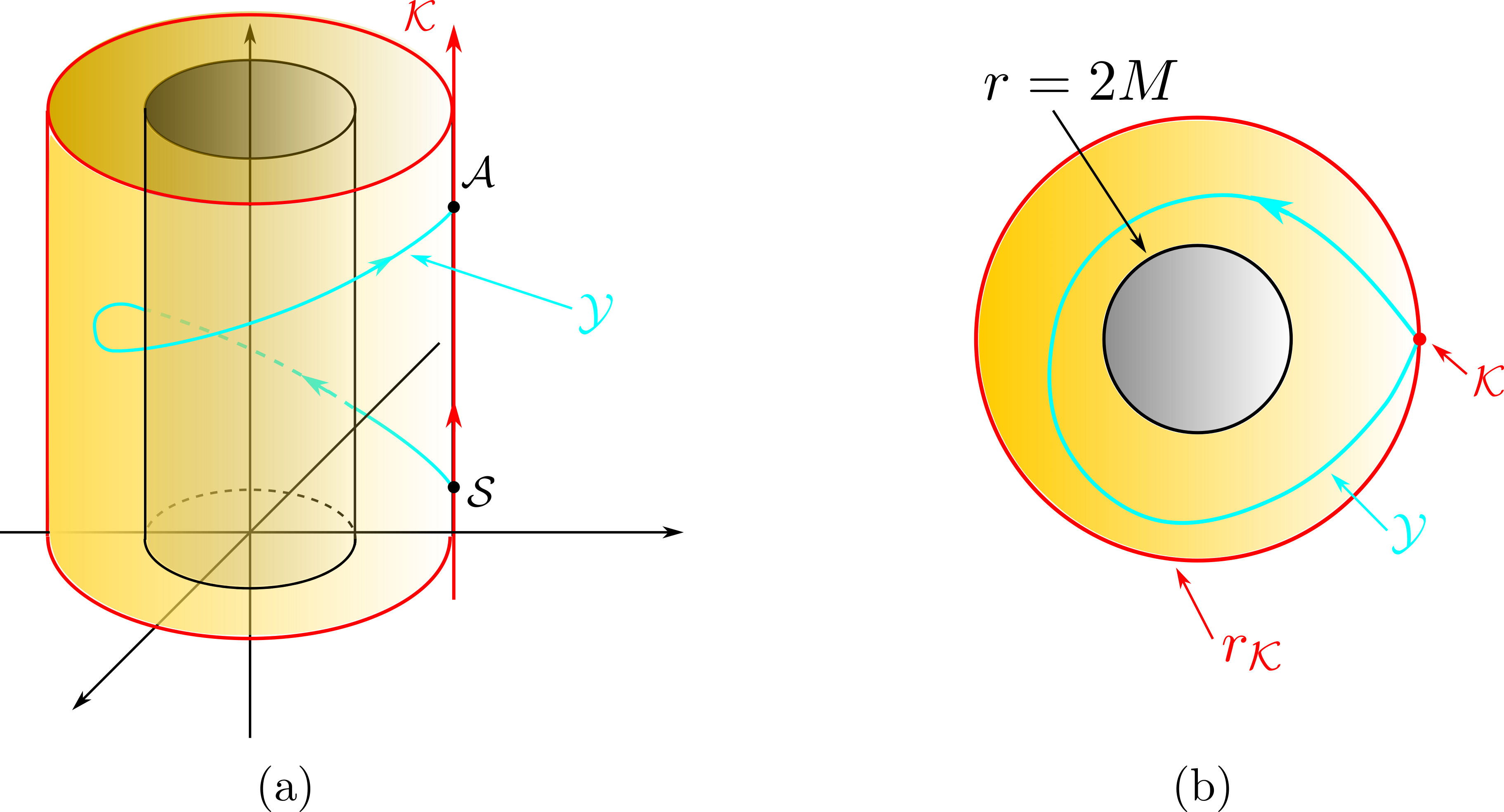}
	\caption{The fountain of youth region in the Schwarzschild spacetime, relative to a static Killing \gls{observer}.}
	\label{fig:observer_K_and_fountain_of_youth}
\end{figure}

For the vacuum Schwarzschild \gls{spacetime}\footnote{Beware: in the literature, it is common to refer to the whole vacuum Schwarzschild spacetime also as the exterior Schwarzschild spacetime (in contrast to the interior, matter-filled one).}, we will work within the usual (symmetry-adapted) spherical coordinate \gls{chart}, $x^\alpha\coloneqq (t,r,\theta,\varphi)$\,, in which the line element is given by
\begin{equation}
	\label{metric}
	ds^2 = -\left(1-\frac{2M}{r}\right)\,dt^2 + \left(1-\frac{2M}{r}\right)^{-1}\,dr^2 + r^2\,d\theta^2 + r^2\sin^2\theta\, d\varphi^2,
\end{equation}
where $M$ is the mass. Such a metric satisfies Einstein's field equations of general relativity, with a vanishing energy-momentum tensor and a vanishing cosmological constant, for the following domain of the coordinate functions: $-\infty < t < \infty, 0 < r < \infty, 0 < \theta , \pi, 0 < \varphi < 2\pi$. This is the single \gls{coordinate_system} we will use throughout the article, restricted, in fact, to the region outside the horizon: $2M< r < \infty$ \footnote{Technically, we have two charts: one for $0<r<2M$\,, and another for $2M < r < \infty$\,.}, though we could easily extend the results to the region inside. The \gls{congruence} of static Killing \glspl{observer} defines a sort of privileged fiducial \gls{reference_frame} (comoving with the aforementioned chart), which is irrotational, rigid (expansionless and shear-free), but non-geodesic (cf. \cite{Sachs1977}). Its proper acceleration is  
\begin{equation}
	a^\alpha = \dfrac{M}{r^2}\delta^\alpha_1\quad \implies \quad a = \dfrac{M}{r^2\sqrt{1-2M/r}}\,.
\end{equation}	
We can choose a corresponding \gls{instantaneous_observer} $\left( (t, r, \theta, \varphi), (dt/d\tau, \boldsymbol{0}) \right)$  whose infinitesimal proper time is, of course, 
		\begin{equation}
			d\tau_\mathcal{K} = \sqrt{\left(1 - \frac{2M}{r_\mathcal{K}}\right)}dt\,. \label{infinitesimal_proper_time_along_K}
		\end{equation}
For such a static \gls{observer}, $\mathcal{K}$ say (cf. Fig. \ref{fig:observer_K_and_fountain_of_youth}), with fixed spatial coordinates $(r=r_\mathcal{K}, \theta=\theta_\mathcal{K}, \varphi=\varphi_\mathcal{K})$\,, and, along it, two arbitrary events $\mathcal{S}:(t=t_\mathcal{S},r=r_\mathcal{S}=r_\mathcal{K},\theta=\theta_\mathcal{S}=\theta_\mathcal{K}, \varphi=\varphi_\mathcal{S}=\varphi_\mathcal{K})$ and $\mathcal{A}:(t=t_\mathcal{A},r=r_\mathcal{A}=r_\mathcal{K},\theta=\theta_\mathcal{A}=\theta_\mathcal{K}, \varphi=\varphi_\mathcal{A}=\varphi_\mathcal{K})$\,, with $\mathcal{A}$ in the future of $\mathcal{S}$\,, i.e., $t_\mathcal{A} > t_\mathcal{S}$. Of course, the corresponding time coordinate interval between such events is simply

\begin{equation}
	\Delta t(\mathcal{S}, \mathcal{A}) = t_\mathcal{A} - t_\mathcal{S}\,,
\end{equation}
and the corresponding proper time interval, along $\mathcal{K}$ [cf. \eqref{infinitesimal_proper_time_along_K}]\,, is
\begin{align}
	\Delta\tau_{\mathcal{K}_{\mathcal{S}\to\mathcal{A}}} &= \sqrt{\left(1-\frac{2M}{r_\mathcal{K}}\right)}\Delta t(\mathcal{S},\mathcal{A})\,. \label{K_proper_time}
\end{align}

Along an arbitrary \gls{observer}, from Eq. \eqref{metric}, the corresponding elapsed infinitesimal proper time may be written as
\begin{equation}
  d\tau^2 = \left( 1-2M/r \right)dt^2 - \left(1 - 2M/r\right)^{-1}dr^2 -r^2\left(d\theta^2 + \sin^2\theta\, d\varphi^2\right)\,.
  \label{infinitesimal_proper_time_along_arbitrary_observer}
\end{equation}
The time coordinate $t$ is well defined within the whole exterior region ($2M<r<\infty$), and so we can also parametrize the mentioned arbitrary observer by a static Killing observer's proper time; thus, by using \eqref{infinitesimal_proper_time_along_K} and \eqref{infinitesimal_proper_time_along_arbitrary_observer}, we have

\begin{equation}
    d\tau^2 = d\tau_{\mathcal{K}}^2\left\{ \dfrac{1-2M/r}{1-2M/r_{\mathcal{K}}} - \dfrac{1}{1-2M/r} \left(\dfrac{dr}{d\tau_{\mathcal{K}}}\right)^2 - r^2\left[ \left( \dfrac{d\theta}{d\tau_{\mathcal{K}}} \right)^2 + \sin^2\theta\left( \dfrac{d\varphi}{d\tau_{\mathcal{K}}} \right)^2 \right]\right\}\,. \label{dtaudtauK}
\end{equation}
Inasmuch as, by definition, an \gls{observer} is timelike, the expression within braces must be positive, i.e., the first term within braces must dominate the last three ones, which are negative definite. However, it is not guaranteed in general that such an expression is definitely greater (or smaller) than 1 and thus we cannot state a priori that an arbitrary observer gets older (or younger) than a static one, $\mathcal{K}$\,.

Nevertheless, we will now show that there is a generous collection of \glspl{observer}, say $\mathcal{Y}$, which will become \textit{younger} than a corresponding static Killing \gls{observer} $\mathcal{K}$\,. To that end, given a static observer $\mathcal{K}$, with radius $r_{\mathcal{K}}$\,, choose a distinct observer $\mathcal{Y}$ which carries out an arbitrary trip not further than $\mathcal{K}$\,, i.e., with $r\le r_{\mathcal{K}}$ everywhere, which implies $(1-2M/r)/(1-2M/r_\mathcal{K})\le 1$\,. We now see that the first term in the braces of Eq. \eqref{dtaudtauK} is already less than 1 and therefore
\begin{equation}
		d\tau_\mathcal{Y}<d\tau_\mathcal{K}\,. \label{dtauYledtauK}
\end{equation}
The total proper time is given by the integral of $d\tau_\mathcal{Y}$ from $\mathcal{S}$ up to $\mathcal{A}$ and thus, provided Eq. \eqref{dtauYledtauK} holds all along, we conclude
\begin{equation}
		\Delta \tau_\mathcal{Y}\leq \Delta\tau_\mathcal{K}\;,
\end{equation}
and the equality holds only for $\mathcal{Y} = \mathcal{K}$ (otherwise the spatial coordinates would vary at some interval and we would have $d\tau_\mathcal{Y}<d\tau_\mathcal{K}$ in this interval). In other words, every arbitrary (non-static) \gls{observer} $\mathcal{Y}$\,, starting at $\mathcal{S}$ and arriving at $\mathcal{A}$\,, that travels everywhere inside a (4-dimensional) cylindrical shell (annulus) $2M<r\leq r_\mathcal{S} = r_\mathcal{A} = r_\mathcal{K}$ in Schwarzschild spacetime returns younger than the corresponding static Killing \gls{observer} at $r_{\mathcal{K}}$: we find a fountain of youth region of spacetime, with respect to the particular observer $\mathcal{K}$ [cf. panel (a) of Fig. \ref{fig:observer_K_and_fountain_of_youth}]! In the spatial projection, the fountain of youth is a spherical shell covering the region $2M<r\leq r_\mathcal{K}$ [cf. panel (b) of Fig. \ref{fig:observer_K_and_fountain_of_youth}].

\FloatBarrier

In this sense, in relation to the static Killing \gls{observer} $\mathcal{K}$\,, going closer to the source indeed will cause our clock to run slower. However, the argument does not hold if the observer wanders outside the fountain of youth, relative to $\mathcal{K}$\,, since we would typically have additional kinetic terms of varying coordinates which would prevent ascertaining the value for the ratio $d\tau/d\tau_{\mathcal{K}}$; therefore, we should be careful when employing this kind of general claim. Below (see Subsec. \ref{subsec:D_and_C}) we will provide a concrete example illustrating that the fountain of youth region is not a fountain of youth relative to general \glspl{observer} which travel outside it.

\begin{figure}[ht]
	\centering
	\includegraphics[scale=0.13]{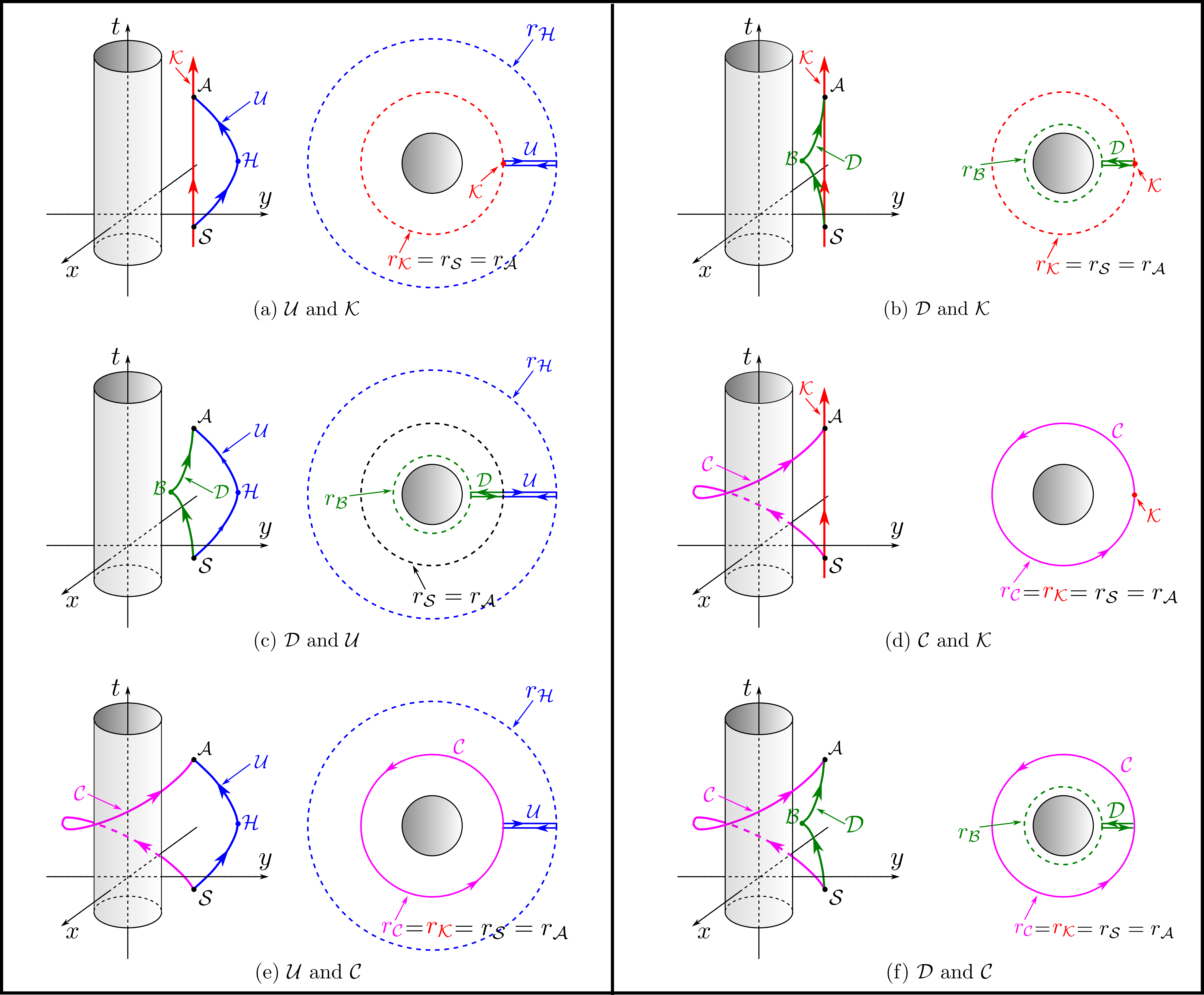}
	\caption{Diagrams for the six couples of twins (\glspl{observer}) considered in the vacuum Schwarzschild \gls{spacetime}; the left and right sides of each panel, from (a) to (f), show, respectively, the corresponding spacetime diagram and the spatial projection onto the equatorial 2-plane $\theta=\pi/2$\,. (a) the couple $\mathcal{U}$ and $\mathcal{K}$\,. (b) the couple $\mathcal{D}$ and $\mathcal{K}$\,. (c) the couple $\mathcal{D}$ and $\mathcal{U}$\,. (d) the couple $\mathcal{C}$ and $\mathcal{K}$\,. (e) the couple $\mathcal{U}$ and $\mathcal{C}$\,. (f) the couple $\mathcal{D}$ and $\mathcal{C}$\,.}. 
 	\label{fig:couples_of_twins}
\end{figure}

To deal with more quantitative results for differential aging, we will consider the four typical \glspl{observer} $\mathcal{K}$, $\mathcal{U}$, $\mathcal{D}$, and $\mathcal{C}$ and some related events, which are listed in Table \ref{tab:observers_and_events} and represented in Fig. \ref{fig:couples_of_twins}. Thus the twins to be dealt with in the comparison of their corresponding proper times are: $(\mathcal{U}, \mathcal{K})$\,, $(\mathcal{D}, \mathcal{K})$\,, $(\mathcal{D}, \mathcal{U})$\,, $(\mathcal{C}, \mathcal{K})$\,, $(\mathcal{U}, \mathcal{C})$\,, and $(\mathcal{D}, \mathcal{C})$\,. It is worth mentioning that Sokolowski \cite{Sokolowski2012} examined the cases involving only observers $\mathcal{K}$, $\mathcal{U}$ and $\mathcal{C}$, illustrating the results in the form of a numerical table.  He stated that $\mathcal{U}$ describes the longest possible curve between the events $\mathcal{S}$ and $\mathcal{A}$ since the geodesic followed by him contains no conjugate points \cite{Hawking1973, Wald1984} in this interval. Later \cite{Sokolowski2017}, he noticed that this result is valid only for the neighborhood of the fiducial curve. We will now characterize all four such \glspl{observer} in more detail.

\begin{table}
	\centering
	\begin{tabular}{|c|c|}
		\hline
		Symbol & Description \\
		\hline
		$\mathcal{K}$ & Static \underline{K}illing observer \\
            $\mathcal{Y}$ & \underline{Y}ounger observer, with respect to $\mathcal{K}$ \\
            $\mathcal{R}$ & Generic \underline{r}adial geodesic observer $[\dot{\varphi}(\tau)\equiv 0]$ \\
		$\mathcal{U}$ & Radial geodesic observer launched \underline{u}pward [$\dot{r}(\mathcal{S})>0$]\,, \\
		& reaching a halt (stop) event $\mathcal{H}$ \\
		$\mathcal{D}$ & Radial geodesic observer, released from rest [$\dot{r}(\mathcal{S})=0$]\,, \\
            & falling initially \underline{d}ownward, reaching a bouncing event $\mathcal{B}$\\
		$\mathcal{C}$ & \underline{C}ircular geodesic observer \\
		$\mathcal{S}$ & \underline{S}tart (departure) event \\
		$\mathcal{A}$ & \underline{A}rrival (``reunion'') event \\
            $\mathcal{P}$ & Generic \underline{p}ause event on the generic radial geodesic observer $[\dot{r}(\mathcal{P})=0]$: \\
            & $\mathcal{H}$ on $\mathcal{U}$\,, or $\mathcal{S}$ and $\mathcal{A}$ on $\mathcal{D}$ \\
		$\mathcal{H}$ & \underline{H}alt (stop) event [$\dot{r}(\mathcal{H})=0$] on $\mathcal{U}$ \\
		$\mathcal{B}$ & \underline{B}ouncing event [$\lim_{\tau\to\tau(\mathcal{B})_+}\dot{r}(\tau) =-\lim_{\tau\to\tau(\mathcal{B})_-}\dot{r}(\tau)$] on $\mathcal{D}$\\
		\hline
	\end{tabular}
	\caption{Symbols and description of relevant \glspl{observer} and events.}
	\label{tab:observers_and_events}
\end{table}

\subsection{Geodesic observers}
\label{subsec:geodesic_observers} 

In a Lorentzian \gls{spacetime} $(\mathcal{M}, g_{\alpha\beta}(x^\mu))$\,, its geodesics can be thought of as the extremal curves of the variational problem associated with the Lagrangian
\begin{equation}
    \label{Lagrangian}
    \mathcal{L}(x^\mu(\tau), \dot{x}^\mu(\tau))=\frac{1}{2}g_{\alpha\beta}(x^\mu(\tau))\,\dot{x}^\alpha(\tau)\,\dot{x}^\beta(\tau)\,,
\end{equation}
where the parameter $\tau$ for the extremal curves turns out to be any affine parameter. 
For timelike geodesics, it can then be chosen, without any loss of generality, as its proper time. Thus, since the Lagrangian itself is a constant of motion, it is apparent, on the geodesics, to be normalized such that
\begin{equation}
	\label{general_Lagrangian_normalization}
     2\mathcal{L}(x^\mu, \dot{x}^\mu)=-1\,.
\end{equation}
Due to the spherical symmetry of \eqref{metric}, we can, again without any loss of generality, restrict our study to equatorial geodesics $\theta\equiv\pi/2$, the associated $x^2\coloneqq\theta$ component of the geodesic equation being identically satisfied. 

The (covariant) canonical momentum associated to coordinate $x^\alpha$ is defined by
\begin{equation}
	\label{definition_of_canonical_momenta}
	p_\alpha \coloneqq \dfrac{\partial\mathcal{L} }{\partial \dot{x}^\alpha} = g_{\alpha\beta}(x^\mu)\,\dot{x}^\beta\,.
\end{equation}
Taking into account \eqref{metric}, since the coordinates $x^0\coloneqq t$ and $x^3\coloneqq\varphi$ are ignorable (cyclic) for the Lagrangian \eqref{Lagrangian}, the corresponding canonical momenta are constants of motion:
\begin{subequations}
\begin{align}
	p_0 &= -\left( 1 - \dfrac{2M}{r} \right)\dot{t}\eqqcolon -E=\text{const}<0\,, \label{generic_p0}\\
	p_3 &= r^2 \sin^2 \theta\,\dot{\varphi}=r^2\,\dot{\varphi} \eqqcolon L=\text{const}\,, \label{generic_p3}
\end{align}
\end{subequations}
with $E$ interpreted as a sort of energy (per mass) of the geodesic particle, whereas $L$ is a sort of its azimuthal angular momentum (also per mass). 
For these equatorial geodesics, we now use \eqref{metric} into \eqref{general_Lagrangian_normalization} to obtain
\begin{equation*}
    \label{normalization_of_Lagrangian}
-\left( 1 - \dfrac{2M}{r}  \right)\dot{t}^2 + \left( 1 - \dfrac{2M}{r}  \right)^{-1}\dot{r}^2 + r^2\,\dot{\varphi}^2=-1 \,,
\end{equation*}
and then \eqref{generic_p0} and \eqref{generic_p3} to get rid of $\dot{t}$ and $\dot{\varphi}$, leading us to
\begin{equation}
    \label{1d_radial_problem}
    \left(\frac{dr}{d\tau}\right)^2 = \left( 1-2M/r \right)\left(-1+\frac{E^2}{1-2M/r}-\frac{L^2}{r^2}\right) \ge 0\,.
\end{equation}
This is the well-known relativistic analog of the reduction of the Keplerian radial problem to a one-dimensional problem \cite{Chandrasekhar1983,Misner1973}; in fact \eqref{1d_radial_problem} can be rewritten as
\begin{equation}
\label{1d_radial_problem_with_effective_potential}
 \left(\frac{dr}{d\tau}\right)^2 = E^2 -V(r)\,,
\end{equation}
where $V(r)$ plays the role of an effective potential, defined by \cite[Sec. 9.3]{Hartle2003} \cite[Sec. 5.4]{Carroll2004}:
\begin{equation}
\label{effective_potential}
V(r; M, L) := \left(1-\frac{2M}{r}\right)\left(1+\frac{L^2}{r^2}\right)\,.
\end{equation}

The  relevant parametric geodesic equations can be cast in the form
\begin{subequations}
	\label{exterior_parametric_geodesics}
	\begin{align}
 \label{eq-geo1a}
		\frac{d\tau}{dr} &= \pm\left[ E^2 - V(r; M, L) \right]^{-1/2}\,, \\
  \label{eq-geo1b}
		\frac{dt}{d\tau} &= \frac{E}{1-2M/r}\,, \\
  \label{eq-geo1c}
		\frac{d\varphi}{d\tau} &= \frac{L}{r^2}\,, \\
  \label{eq-geo1d}
		\frac{dt}{dr} &= \pm E\left(1-\frac{2M}{r}\right)^{-1}\left[  E^2 - V(r; M, L) \right]^{-1/2}\,.
	\end{align}
\end{subequations}

\subsubsection{Generic \emph{radial} geodesic observers $\mathcal{R}$}
\label{subsubsec:radial_geodesic_observers}

For radial geodesics, $L=0$,  \eqref{exterior_parametric_geodesics} simplifies to
\begin{subequations}
	\label{radial_parametric_geodesics}
	\begin{align}
		\frac{d\tau}{dr} &= \pm\left( E_\mathcal{R}^2 - 1 + \frac{2M}{r} \right)^{-1/2}\,, \label{ext_rad_param_dtaudr} \\
		\frac{dt}{d\tau} &= \frac{E_\mathcal{R}}{1-2M/r}\,, \label{ext_rad_param_dott}\\
		\frac{d\varphi}{d\tau} &= 0\,, \label{ext_rad_param_dotvarphi} \\
		\frac{dt}{dr} &= \pm E_\mathcal{R}\left(1-\frac{2M}{r}\right)^{-1}\left(  E_\mathcal{R}^2 -1 + \frac{2M}{r} \right)^{-1/2}\,, \label{ext_rad_param_dtdr}
	\end{align}
\end{subequations}
where now the energy parameter is denoted by $E_\mathcal{R}$\,.

The only branches (arcs) of radial geodesic \glspl{observer} we will consider in this subsubsection are those that are infalling; they may be thought of as constituting either the infalling branch $\mathcal{U}_{\mathcal{H}\to\mathcal{A}}$ of observer $\mathcal{U}$, or the infalling branch $\mathcal{D}_{\mathcal{S}\to\mathcal{B}}$ of {observer} $\mathcal{D}$\,. In any case, there is an event $\mathcal{P}$ at which the radial {observer} is momentarily paused (at rest) with respect to a corresponding static Killing {observer}, $\dot{r}(\mathcal{P})=0$\,, viz.: for $\mathcal{U}_{\mathcal{H}\to\mathcal{A}}$\,, the halt event $\mathcal{H}$, or, for $\mathcal{D}_{\mathcal{S}\to\mathcal{B}}$\,, the start event $\mathcal{S}$ (cf. Fig.~\ref{fig:couples_of_twins})\,.

At such particular pause events $\mathcal{P}$ (on $\mathcal{R}$) we thus have, from \eqref{ext_rad_param_dtaudr}, the following constraint
\begin{equation}
	0 < E_\mathcal{R} = E_\mathcal{P} := \sqrt{1-2M/r_{\mathcal{P}}} < 1  \iff r_\mathcal{P} = \dfrac{2M}{1-E_\mathcal{P}^2}\,. \label{constraint_between_E_and_r_R}
\end{equation}

Equations \eqref{ext_rad_param_dtaudr} and \eqref{ext_rad_param_dtdr} can be integrated into a closed but complicated form. However, a much simpler procedure, following the lead of Misner, Thorne, and Wheeler \cite{Misner1973} (cf. also Chandrasekhar \cite{Chandrasekhar1983}), consists in  defining a new coordinate $\eta$ as
\begin{equation}
    \label{ext_rad_eta_r_rm}
    0\leq\eta:= \arccos\left(\frac{2r}{r_{\mathcal{P}}}-1\right)\leq \pi\,,
\end{equation}
so $\eta=0$ for  $r=r_{\mathcal{P}}$ (and, were we to extend the arc up to the singularity $r=0$, then $\eta=\pi$)\,. Thus
\begin{equation}
    \label{ext_rad_Chandrasekhar_coordinates}
    r = \frac{r_{\mathcal{P}}}{2}(1+\cos \eta) = r_{\mathcal{P}}\cos^2({\eta}/
{2})\,.
\end{equation}
It is also convenient to define the value $\eta=\eta_h$ at the associated static black hole horizon radius  $r_h:=2M$, 
such that
\begin{equation}
    \label{ext_rad_eta_E}
    \eta_h=2\arcsin E_\mathcal{P}>0\,.
\end{equation}
In terms of $\eta$, it is not difficult to show that
\begin{equation}
    \label{ext_rad_dtaudeta}
    \frac{d\tau}{d\eta}=\sqrt{\frac{r_{\mathcal{P}}^3}{2M}}\cos^2({\eta}/{2})\,,
\end{equation}
and
\begin{equation}
   \label{ext_rad_dtdeta}
   \dfrac{dt}{d\eta} = E_\mathcal{P}\sqrt{\frac{r_{\mathcal{P}}^3}{2M}}\,
    \frac{\cos^4({\eta}/{2})}{\cos^2({\eta}/{2})-\cos^2({\eta_h}/{2})}\,.
\end{equation}

\noindent Integration of \eqref{ext_rad_dtaudeta} and \eqref{ext_rad_dtdeta}, {from $\mathcal{P}: (\eta_\mathcal{P} = \eta(r_\mathcal{P})=0)$ to a generic event (to the future of $\mathcal{P}$) $\mathcal{E}: (\eta_\mathcal{E} = \eta(r_\mathcal{E}) >0)$} , gives, respectively, 
\begin{equation}
    \label{ext_rad_tau(eta)}
    \Delta\tau_{\mathcal{R}_{\mathcal{P}\to\mathcal{E}}}{(M, r_\mathcal{P}, r_\mathcal{E})} =  \sqrt{\frac{r_{\mathcal{P}}^3}{8M}}\,(\eta_\mathcal{E} +\sin \eta_\mathcal{E})\,,
\end{equation}
and
\begin{equation}
    \label{ext_rad_t(eta)}
     \Delta t_{\mathcal{R}_{\mathcal{P}\to\mathcal{E}}}{(M, r_\mathcal{P}, r_\mathcal{E})} = E_\mathcal{P}\sqrt{\frac{r_{\mathcal{P}}^3}{2M}}F_1(\eta_\mathcal{E}, E_\mathcal{P})+2MF_2(\eta_\mathcal{E}, \eta_h)\,,
\end{equation}

\noindent where 
\begin{subequations}
    \label{ext_rad_F(eta)}
\begin{align}
     F_1(\eta_\mathcal{E}, E_\mathcal{P}) &:= \frac{1}{2}(\eta_\mathcal{E} +\sin\eta_\mathcal{E})+\eta_\mathcal{E}(1-E_\mathcal{P}^2)\,, \\
     F_2(\eta_\mathcal{E}, \eta_h) &:= \ln\left[\frac{\tan(\eta_h/2)+\tan(\eta_\mathcal{E}/2)}{\tan(\eta_h/2)-\tan(\eta_\mathcal{E}/2)}\right]\,.
\end{align}
\end{subequations}

\subsubsection{Generic \emph{circular} geodesic observers $\mathcal{C}$}
\label{subsubsec:circular_geodesic_observers}

{The roots of $dV/dr=0$ or, equivalently, 
\begin{equation}
    Mr_\mathcal{C}^2 - L_\mathcal{C}^2r_\mathcal{C} + 3ML_\mathcal{C}^2 = 0\,, \label{vanishing_derivative_eff_pot}
\end{equation}    
i.e.,
\begin{equation}
\label{circ_rc}
r_{\mathcal{C}\pm} = \frac{L_\mathcal{C}^2}{2M}\left(1\pm\sqrt{1-\frac{12M^2}{L_\mathcal{C}^2}}\right)\,,
\end{equation}
are potential candidates for the radii of circular geodesic orbits; for them to correspond to actual circular geodesics, they have to be greater than $2M$. This is satisfied if and only if $L_\mathcal{C}^2\ge 12M^2$, implying, of course, the reality and positivity of the roots. The larger root, $r_{\mathcal{C}+}$\,, corresponds to a stable circular orbit (local minimum of $V$) and the smaller root, $r_{\mathcal{C}-}$\,, corresponds to an unstable circular orbit (local maximum of $V$). For the minimum value $L_\mathcal{C}^2/M^2=12$ we have $r_{\mathcal{C}\pm}=6M$, which is the infimum for the radius of a stable circular orbit: $r_{\mathcal{C}+}\ge 6M$. The infimum for the radius of an unstable circular orbit corresponds to $L_\mathcal{C}\rightarrow \infty$; in this case, $12M^2/L_\mathcal{C}^2\rightarrow 0$ and $\sqrt{1-12M^2/L_\mathcal{C}^2}\approx 1-6M^2/L^2$, so $r_{\mathcal{C}-} > 3M$. Thus, there exist unstable circular geodesic orbits for $3M < r_{\mathcal{C}-} \leq 6M$, while for $6M < r_{\mathcal{C}+} < \infty$, we have stable circular geodesic orbits. }

Let us now find the period of the orbit in terms of proper and coordinate time.  Solving \eqref{vanishing_derivative_eff_pot} for $L_\mathcal{C}^2$\,, we find the relation between the angular momentum for any circular geodesic and its radius:
\begin{equation}
\label{circ_L2}
L_\mathcal{C}^2 = \frac{Mr_{\mathcal{C}}}{1-3M/r_\mathcal{C}}\,.
\end{equation}
In such a case, since the corresponding energy satisfies $E_\mathcal{C}^2=V(r_\mathcal{C})$, we find, after substituting \eqref{circ_L2} into \eqref{effective_potential}, the relation between the energy of any circular geodesic and its radius:
\begin{equation}
\label{circ_E2}
E_\mathcal{C} = \frac{1-2M/r_\mathcal{C}}{\sqrt{1-3M/r_\mathcal{C}}}\,.
\end{equation}
From \eqref{eq-geo1b} and \eqref{eq-geo1c} we have
\begin{equation}
\label{circ_dtdvarphi_EL}
\frac{dt}{d\varphi}=\frac{E_\mathcal{C}\,r_\mathcal{C}^2}{L_\mathcal{C}(1-2M/r_\mathcal{C})}\,.
\end{equation}
Substituting for \eqref{circ_L2} and \eqref{circ_E2} in the above equation we obtain
\begin{equation}
\label{circ_dtdvarphi_rc}
\frac{dt}{d\varphi} = {\pm}\sqrt{\frac{r_\mathcal{C}^3}{M}}\,.
\end{equation}
The elapsed coordinate time for a round trip connecting a starting event $\mathcal{S}$ to an arrival event $\mathcal{A}$ is then given by
\begin{equation}
\label{circ_deltatc}
\Delta t_{\mathcal{C}_{\mathcal{S}\to\mathcal{A}}}(M, r_\mathcal{C}) = 2\pi\sqrt{\frac{r_\mathcal{C}^3}{M}}\,.
\end{equation}
The proper time can be obtained immediately from \eqref{eq-geo1b}:
\begin{equation}
\label{circ_deltatauc}
\Delta\tau_{\mathcal{C}_{\mathcal{S}\to\mathcal{A}}}(M, r_\mathcal{C}) =  \frac{1-2M/r_\mathcal{C}}{E_\mathcal{C}}\,\Delta t_{\mathcal{C}_{\mathcal{S}\to\mathcal{A}}}(M,r_\mathcal{C}) = 2\pi r_\mathcal{C}\sqrt{\left(1-\frac{3M}{r_\mathcal{C}}\right)\frac{r_\mathcal{C}}{M}}\,.
\end{equation}

\section{Proper time ratios for twins}
\label{sec:comparison_proper_times}

The results in subsections \ref{subsec:U_and_K}, \ref{subsec:C_and_K} and \ref{subsec:U_and_C} are partially included in \cite{Sokolowski2012}  but he did not present any plots for the proper time ratios, just some numerical tables.

\subsection{Observers $\mathcal{U}$ and $\mathcal{K}$}
\label{subsec:U_and_K}

 Here we aim to solve the following problem: the traveling geodesic \gls{observer} $\mathcal{U}$ (cf.\ Fig.~\ref{fig:couples_of_twins}) is launched radially upward from the starting event $\mathcal{S}: (t=t_{\mathcal{S}}, r = r_{\mathcal{S}})$\,, travels until a halting event $\mathcal{H}: (t=t_{\mathcal{H}}, r=r_{\mathcal{H}})$ and returns, falling back to its initial position  $r=r_{\mathcal{S}}$ at the arrival event $\mathcal{A}: (t = t_{\mathcal{A}}, r = r_{\mathcal{A}} = r_{\mathcal{S}})$. Let $\Delta\tau_{\mathcal{U}_{\mathcal{S}\to\mathcal{A}}}$ denote the journey's total proper time registered by the {observer} $\mathcal{U}$. Let $\Delta\tau_{\mathcal{K}_{\mathcal{S}\to\mathcal{A}}}$ denote the corresponding  proper 
time registered by a static {observer} $\mathcal{K}$ that remains at the position $r_{\mathcal{K}}=r_{\mathcal{S}} = r_{\mathcal{A}}$. We want to 
find the relation $\Delta\tau_{\mathcal{U}_{\mathcal{S}\to\mathcal{A}}}/\Delta\tau_{\mathcal{K}_{\mathcal{S}\to\mathcal{A}}}$. In what follows, we will determine the proper time   
 $\Delta\tau_{\mathcal{U}_{\mathcal{H}\to\mathcal{A}}}$ associated with the partial trip of the {observer} $\mathcal{U}$ from $\mathcal{H}$ (where it is momentarily at rest with respect to the corresponding static Killing {instantaneous observer})
until it reaches $\mathcal{A}$. By symmetry, the proper time associated with the upward trip is the same, so the total elapsed proper time is
\begin{equation}
\label{DeltatauU_StoA}
\Delta\tau_{\mathcal{U}_{\mathcal{S}\to\mathcal{A}}} = 2\Delta\tau_{\mathcal{U}_{\mathcal{H}\to\mathcal{A}}}\,.   
\end{equation}

Thus, using  \eqref{ext_rad_tau(eta)} into \eqref{DeltatauU_StoA}, with $\mathcal{R}_{\mathcal{P}\to\mathcal{E}}=\mathcal{U}_{\mathcal{H}\to\mathcal{A}}$\,, $r_\mathcal{P}=r_\mathcal{H}$, and  $r_\mathcal{E} = r_{\mathcal{A}} = r_{\mathcal{S}}$\,,
we get
\begin{subequations}
\label{DeltatauU_total}
\begin{equation}
\label{ext_Deltatau_U}
    \Delta\tau_{\mathcal{U}_{\mathcal{S}\to\mathcal{A}}}(M, r_{\mathcal{H}}, r_{\mathcal{S}}) = \sqrt{\dfrac{r_{\mathcal{H}}^3}{2M}}(\eta_{\mathcal{S}} + \sin\eta_{\mathcal{S}})\,,
\end{equation}
where  $\eta_{\mathcal{S}}$ is given by
\begin{equation}
    \label{ext_U_etaS}
    \eta_{\mathcal{S}} = \arccos\left(\dfrac{2r_{\mathcal{S}}}{r_{\mathcal{H}}}-1\right)\,.
\end{equation}
\end{subequations}
We have chosen to express such a proper time as an explicit function of the three variables $(M, r_\mathcal{H}, r_\mathcal{S}$)\,.

The corresponding elapsed coordinate time is (cf.\ \eqref{ext_rad_t(eta)} and \eqref{ext_rad_F(eta)})
\begin{subequations}
\label{DeltatU_total}
\begin{align}
    \label{eq-coordinate_time_U}
    \Delta t_{\mathcal{U}_{\mathcal{S}\to\mathcal{A}}}(M, r_\mathcal{H}, r_\mathcal{S}) &= 2\Delta t_{\mathcal{U}_{\mathcal{H}\to\mathcal{A}}}(M, r_\mathcal{H}, r_\mathcal{S}) \nonumber \\
    &= 2E_\mathcal{H}\sqrt{\dfrac{r_{\mathcal{H}}^3}{2M}}F_1(\eta_{\mathcal{S}},E_\mathcal{H}) + 4MF_2(\eta_{\mathcal{S}},\eta_h)\,,
\end{align}
where $E_\mathcal{H}$ and $\eta_H$ are given by
\begin{align}
    E_\mathcal{H} &= \sqrt{1 - 2M/r_\mathcal{H}}\,, \\
    \eta_h &= 2\arcsin E_\mathcal{H}\,. 
\end{align}
\end{subequations}
Let us now consider the static {observer} $\mathcal{K}$ at $r=r_{\mathcal{K}}=r_{\mathcal{S}}= r_\mathcal{A}$. Its proper time associated with the round trip of the {observer} $\mathcal{U}$ is given by [cf.~\eqref{K_proper_time}]

\begin{equation}
   \label{ext_rad_deltatauk}
    {\Delta\tau_{\mathcal{K}_{\mathcal{S}\to\mathcal{A}}}(M, r_\mathcal{H}, r_\mathcal{S}) = \sqrt{1-\frac{2M}{r_{\mathcal{S}}}}\Delta t_{\mathcal{U}_{\mathcal{S}\to\mathcal{A}}}(M, r_\mathcal{H}, r_\mathcal{S})\,,}
\end{equation}
where, since both {observers} start and arrive at the same events, we used $\Delta t(\mathcal{S}, \mathcal{A}) = \Delta t_{\mathcal{U}_{\mathcal{S}\to\mathcal{A}}}$\,.

We want to compare the proper times $\Delta\tau_{\mathcal{U}_{\mathcal{S}\to\mathcal{A}}}$ and $\Delta\tau_{\mathcal{K}_{\mathcal{S}\to\mathcal{A}}}$. From the above equations, we have

\begin{equation}   \label{ext_ratio_U_over_K}
    {R_{\mathcal{U}/\mathcal{K}}(M, r_\mathcal{H}, r_\mathcal{S}):=}\frac{\Delta\tau_{\mathcal{U}_{{\mathcal{S}\to\mathcal{A}}}}{(M, r_\mathcal{H}, r_\mathcal{S})}}{\Delta\tau_{\mathcal{K}_{{\mathcal{S}\to\mathcal{A}}}}{(M, r_\mathcal{H}, r_\mathcal{S})}} \,.
    \end{equation}

\begin{figure}[ht]
\centering
\includegraphics[width=12cm]{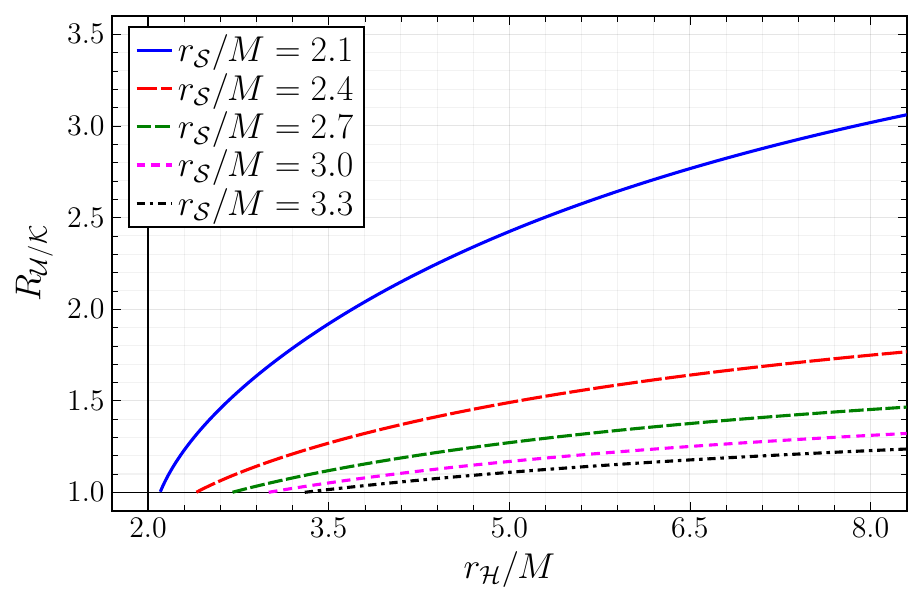}
\caption{Proper time ratio $R_{\mathcal{U}/\mathcal{K}}:=\Delta\tau_\mathcal{U}/\Delta\tau_{\mathcal{K}}$ \emph{versus} the halt radius $r_\mathcal{H}/M$ for distinct values of the start radius $r_\mathcal{S}/M=r_\mathcal{A}/M=r_\mathcal{K}/M$\,.}
\label{fig:ratio_U_over_K}
\end{figure}

We present in Fig. \ref{fig:ratio_U_over_K}  the ratio $R_{\mathcal{U}/\mathcal{K}}$ in terms of $r_\mathcal{H}$ for different values of  $r_\mathcal{S}$.
The graph shows  that $\Delta\tau_{\mathcal{U}}>\Delta\tau_{\mathcal{K}}$ in all cases. In other words, the geodesic traveling \gls{observer} $\mathcal{U}$ comes back older than the static one $\mathcal{K}$.  Larger values of $r_\mathcal{H}$ do not contribute considerably to increasing the proper time ratio since the gravitational field effects become weaker. 
When $r_\mathcal{H} \rightarrow r_\mathcal{S}$, $R_{\mathcal{U}/\mathcal{K}}\rightarrow 1$ as expected. 

\FloatBarrier

It is worthwhile to mention the behavior of some quantities under the following transformation
\begin{align}
    M &\to \lambda M\,,\\
    r_\mathcal{S} &\to \lambda r_\mathcal{S}\,, \\
    r_\mathcal{H} &\to \lambda r_\mathcal{H}\,.
\end{align}
Concretely, the individual proper times $\Delta\tau_\mathcal{U}$ and $\Delta\tau_\mathcal{K}$ are not invariant (nor even homogeneous of any degree) but their ratio $R_{\mathcal{U}/\mathcal{K}}$ is invariant (equivalently, homogeneous of degree zero); in other words, though the ratio given in Eq. \eqref{ext_ratio_U_over_K} is indeed a function of $M, r_\mathcal{H}$ and $r_\mathcal{S}$ \,, it is so only through the ``normalized radii'' $r_\mathcal{S}/M$ and $r_\mathcal{H}/M$\,: if we double the mass, for instance, and simultaneously double both $r_\mathcal{S}$ and $r_\mathcal{H}$\,, that ratio does not change.

The results of this subsection are also partially included in \cite{Gron2011}, although such authors chose to plot the ratio of proper times as a function of $r_\mathcal{K}/M$ instead of $r_\mathcal{H}/M$. They also deal with the ``twin paradox'' in two other spacetimes: 
Minkowski for a ``uniformly accelerated'' (Rindler) {observer}, and Robertson-Walker for a Hubble {observer} and a radially ``accelerated'' one.   

\subsection{Observers $\mathcal{D}$ and $\mathcal{K}$}
\label{subsec:D_and_K}

Let us consider now two other \glspl{observer}: one of them is, as usual, the same kind of fiducial static Killing {observer} $\mathcal{K}$, at $r=r_\mathcal{K}$\,, and another one is the piecewise smooth radial geodesic {observer} $\mathcal{D}$ (cf. Fig.~\ref{fig:couples_of_twins}). More specifically, they start from a common event $\mathcal{S}: (t=t_\mathcal{S}, r=r_\mathcal{S})$, where $\mathcal{D}$ is released from rest (with respect to $\mathcal{K}$), is thus attracted downward and, when reaching an event $\mathcal{B}: (t=t_\mathcal{B}, r=r_\mathcal{B})$\,, bounces in a perfectly elastic way (i.e., $\lim_{\tau\to\tau_\mathcal{B}+}\dot{r}(\tau) = 
-\lim_{\tau\to\tau_\mathcal{B}-}\dot{r}(\tau)$), and again meets $\mathcal{K}$ at an arrival event $\mathcal{A}: (t=t_\mathcal{A},r=r_\mathcal{A})$\,; of course, $r_\mathcal{S} = r_\mathcal{A} = r_\mathcal{K}$\,.

The geodesic arc now, along $\mathcal{D}$\,, from $\mathcal{S}$ to $\mathcal{B}$ is equivalent to the geodesic arc, from the former subsection \ref{subsec:U_and_K}, along $\mathcal{U}$\,, from $\mathcal{H}$ to $\mathcal{A}$\,. Thus the relevant proper times are the total elapsed proper time, along $\mathcal{D}$, from $\mathcal{S}$ to $\mathcal{A}$, parameterized in terms of $M$\,, $r_\mathcal{S}$\,, and $r_\mathcal{B}$ [cf. \eqref{ext_rad_tau(eta)}]:
\begin{subequations}
\begin{align}
\label{eq-proper-time-down}
\Delta\tau_{\mathcal{D}_{\mathcal{S}\to\mathcal{A}}}(M, r_\mathcal{S}, r_\mathcal{B}) &= 2\Delta\tau_{\mathcal{D}_{\mathcal{S}\to\mathcal{B}}}(M, r_\mathcal{S}, r_\mathcal{B}) \nonumber \\
      &= \sqrt{\dfrac{r_\mathcal{S}^3}{2M}}(\eta_\mathcal{B} + \sin\eta_\mathcal{B})\,,
\end{align}
with
\begin{equation}
	\eta_\mathcal{B} = \textrm{arccos}\left( \dfrac{2r_\mathcal{B}}{r_\mathcal{S}} - 1 \right)\,,
\end{equation}
\end{subequations}
and the total elapsed proper time, along $\mathcal{K}$, from $\mathcal{S}$ to $\mathcal{A}$, parameterized in terms of the same variables [cf. \eqref{K_proper_time}]:
\begin{equation}
	\Delta\tau_{\mathcal{K}_{\mathcal{S}\to\mathcal{A}}}(M, r_\mathcal{S}, r_\mathcal{B}) = \sqrt{1 - \dfrac{2M}{r_\mathcal{B}}}
 \Delta t(\mathcal{S},\mathcal{A})\,,
\end{equation}
with
\begin{subequations}
        \begin{align}
        \label{eq-coordinate_time_D}
          \Delta t(\mathcal{S},\mathcal{A}) = \Delta t_{\mathcal{D}_{\mathcal{S}\to\mathcal{A}}}(M, r_\mathcal{S}, r_\mathcal{B}) &= 2E_\mathcal{S}\sqrt{\dfrac{r_\mathcal{S}^3}{2M}} F_1(\eta_\mathcal{B}, E_\mathcal{S}) + 4M F_2(\eta_\mathcal{B}, \eta_h)\,, \\
               F_1(\eta_\mathcal{B}, E_\mathcal{S}) &= \frac{1}{2}(\eta_\mathcal{B} +\sin\eta_\mathcal{B})+\eta_\mathcal{B}(1-E_\mathcal{S}^2)\,, \\
     F_2(\eta_\mathcal{B}, \eta_h) &= \ln\left[\frac{\tan(\eta_h/2)+\tan(\eta_\mathcal{B}/2)}{\tan(\eta_h/2)-\tan(\eta_\mathcal{B}/2)}\right]\,, \\
	E_\mathcal{S} &= \sqrt{1 - 2M/r_\mathcal{S}}\,, \\
  	\eta_h &= 2\,\text{arcsin}\,E_\mathcal{S}\,.
  \end{align}
\end{subequations}     

\begin{figure}[!ht]
\centering
\includegraphics[width=12cm]{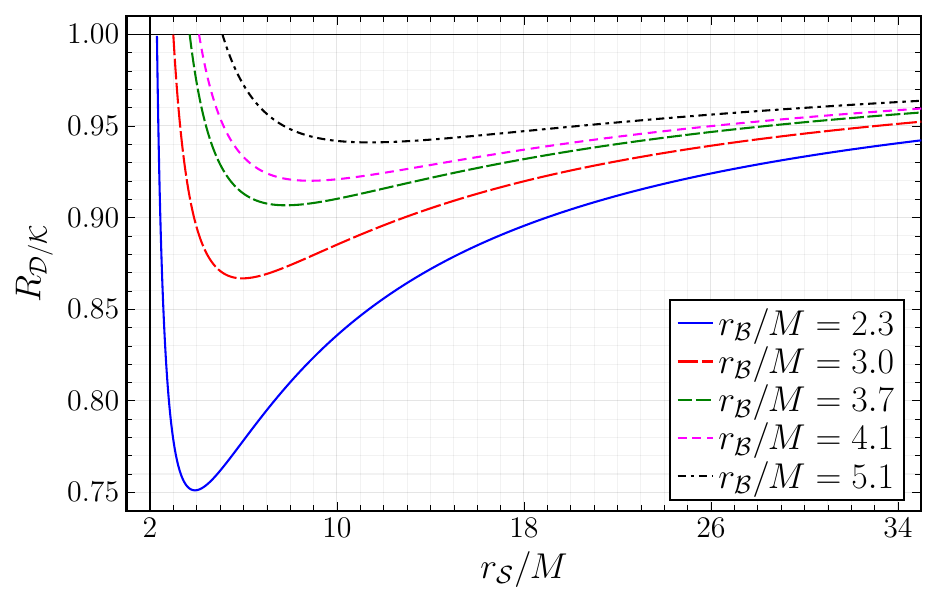}
\caption{{Proper time ratio $R_{\mathcal{D}/\mathcal{K}}\coloneqq \Delta\tau_\mathcal{D}/ \Delta\tau_\mathcal{K}$ \emph{versus} the start radius $r_\mathcal{S}/M = r_\mathcal{A}/M = r_\mathcal{K}/M$ for distinct values of the bouncing radius $r_\mathcal{B}/M$\,.}}
\label{fig:ratio_D_over_K}
\end{figure}
\FloatBarrier

We present proper time ratios $R_{\mathcal{D}/\mathcal{K}} := \Delta \tau_{\mathcal{D}}/\Delta\tau_{\mathcal{K}}$ in terms of $r_{\mathcal{S}}/M$ in Fig. \ref{fig:ratio_D_over_K}, for  different values of $r_{\mathcal{B}}$.

We note that
\begin{itemize}
    \item for any $r_{\mathcal{S}}\ge r_\mathcal{B}$\,, we have  $R_{\mathcal{D}/\mathcal{K}}\le 1$. Indeed, this is expected by our qualitative argument presented at the previous section. The observer $\mathcal{D}$ is a $\mathcal{Y}$-type observer going into the fountain of youth region, therefore returning younger than the static observer $\mathcal{K}$.
    \item for any $2M < r_\mathcal{B} \le r_\mathcal{S}$\,, there is a radius $r{_\mathcal{S}}$ corresponding to a minimum of $R_{\mathcal{D}/\mathcal{K}}$\,.
\end{itemize}

\subsection{Observers $\mathcal{D}$ and $\mathcal{U}$}
\label{subsec:D_and_U}

\begin{figure}[ht]
\centering
\includegraphics[width=12cm]{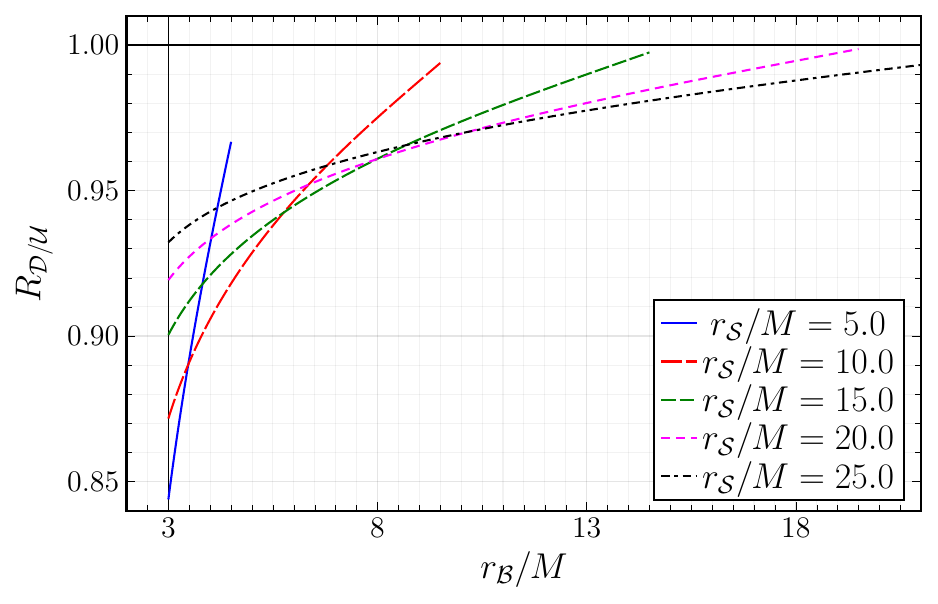}
\caption{
{Proper time ratio $R_{\mathcal{D}/\mathcal{U}}:=\Delta\tau_\mathcal{D}/\Delta\tau_{\mathcal{U}}$ \emph{versus} the bounce radius $r_\mathcal{B}/M$ for distinct values of the start radius $r_\mathcal{S}/M=r_\mathcal{A}/M$\,.}
}
\label{fig:ratio_D_over_U}
\end{figure}

Let us suppose that at an event $\mathcal{S}$, at the radius $r=r_{\mathcal{S}}$ an \gls{observer} $\mathcal{D}$ is released downward and an \gls{observer} $\mathcal{U}$ is vertically launched upward (cf.~Fig.~\ref{fig:couples_of_twins}). To ensure that $\mathcal{D}$ and $\mathcal{U}$ meet again at the same event $\mathcal{S}$ at $r=r_{\mathcal{S}}$\,, observer $\mathcal{D}$ must be bounced, at a radius $r=r_{\mathcal{B}}$, and we ought to select appropriate combinations for the values of the parameters 
$r_{\mathcal{S}}$, $r_{\mathcal{B}}$ and $r_{\mathcal{H}}$, such that the round trip coordinate times satisfy $\Delta t_{\mathcal{U}} =\Delta t_{\mathcal{D}}$ (cf. \eqref{eq-coordinate_time_U} and 
\eqref{eq-coordinate_time_D}). The proper time ratio $R_{\mathcal{D}/\mathcal{U}} :=\Delta \tau_{\mathcal{D}}/\Delta \tau_{\mathcal{U}}$ is shown in Fig. \ref{fig:ratio_D_over_U}.

\FloatBarrier

\subsection{Observers $\mathcal{C}$ and $\mathcal{K}$}
\label{subsec:C_and_K}

\begin{figure}[ht]
\centering
\includegraphics[width=10cm]{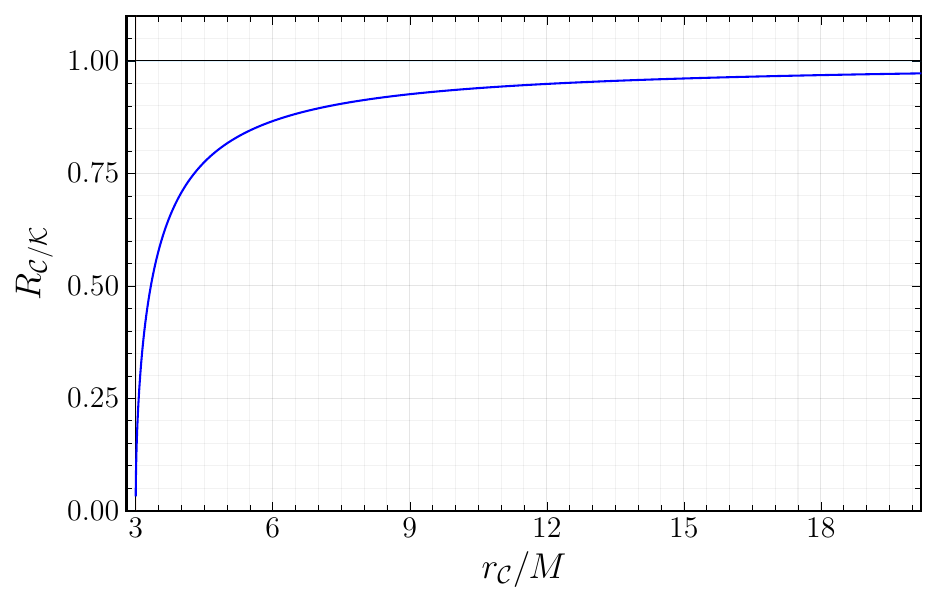}
\caption{
{Proper time ratio $R_{\mathcal{C}/\mathcal{K}}\coloneqq \Delta\tau_\mathcal{C}/ \Delta\tau_\mathcal{K}$ \emph{versus} the circular radius $r_\mathcal{C}/M = r_\mathcal{K}/M =r_\mathcal{S}/M = r_\mathcal{A}/M$\,.}}
\label{fig:ratio_C_over_K}
\end{figure}

Now we consider a circular geodesic \gls{observer} $\mathcal{C}$ at the radius of a comoving static Killing \gls{observer} $\mathcal{K}$ (cf.~Fig.~\ref{fig:couples_of_twins}).
The corresponding elapsed proper time  for $\mathcal{K}$ is given by
\begin{equation}
    \label{circ_deltatauk}
    \Delta\tau_{\mathcal{K}_{\mathcal{S}\to\mathcal{A}}}(M, r_\mathcal{C}) =\sqrt{1-\frac{2M}{r_{\mathcal{C}}}}\,\Delta t_{\mathcal{C}_{\mathcal{S}\to\mathcal{A}}}(M,r_\mathcal{C})\,.
\end{equation}
{Using \eqref{circ_deltatauc},} we now find the ratio
\begin{equation}
    \label{circ_deltataucdeltatauk}
R_{\mathcal{C}/\mathcal{K}}(M, r_\mathcal{C}) :=\frac{\Delta\tau_{\mathcal{C}_{\mathcal{S}\to\mathcal{A}}}(M,r_\mathcal{C})}{\Delta\tau_{\mathcal{K}_{\mathcal{S}\to\mathcal{A}}}(M,r_\mathcal{C})}=\sqrt{\frac{1-3M/r_{\mathcal{C}}}{1-2M/r_{\mathcal{C}}}}<1\,.
\end{equation}
Therefore, {observer} $\mathcal{C}$ always returns younger than $\mathcal{K}$, i.e., the circular geodesic observer ages less than the static accelerated one. This qualitative result is also expected by the argument presented in the previous section since $\mathcal{C}$ is a $\mathcal{Y}$-type observer. The behavior of the ratio $R_{\mathcal{C}/\mathcal{K}}$  in terms of $r_{\mathcal{C}}/M$ is shown in  Fig.~\ref{fig:ratio_C_over_K}.

\FloatBarrier

\subsection{Observers $\mathcal{U}$ and $\mathcal{C}$}
\label{subsec:U_and_C}

Let us consider now an \gls{observer} $\mathcal{C}$ in a circular geodesic orbit. Starting from a common event $\mathcal{S}$, another \gls{observer} $\mathcal{U}$ is launched radially upwards in such a way that $\mathcal{C}$ and $\mathcal{U}$ meet again exactly at the event $\mathcal{A}$ corresponding to a complete orbit (cf.~Fig.~\ref{fig:couples_of_twins}). 

To ensure that $\mathcal{C}$ and $\mathcal{U}$ meet again at the end of their trips, we need to adjust $r_{\mathcal{C}}$ and $r_{\mathcal{H}}$ accordingly, so that the elapsed coordinate times for the round trips of $\mathcal{C}$ and $\mathcal{U}$  coincide. The coordinate time associated to one complete orbit  of $\mathcal{C}$ is given, according to  \eqref{circ_deltatc} 
by
\begin{equation}
\label{circ_deltatc2}
\Delta t_{\mathcal{C}_{\mathcal{S}\to\mathcal{A}}}(M, r_\mathcal{C}) = 2\pi\sqrt{\frac{r_{\mathcal{C}}^3}{M}}\,,
\end{equation}
and the corresponding coordinate time for $\mathcal{U}$ to go up and return is, according to  \eqref{eq-coordinate_time_U}, given by
\begin{equation}
    \label{ext_Deltat_K}
    \Delta t_{\mathcal{U}_{\mathcal{S}\to\mathcal{A}}} (M,r_\mathcal{H},r_\mathcal{S})= 2E_\mathcal{H}\sqrt{\dfrac{r_{\mathcal{H}}^3}{2M}}F_1(\eta_{\mathcal{S}},E_\mathcal{H}) + 4MF_2(\eta_{\mathcal{S}},\eta_h)\,.
\end{equation}
where $F_1$ and $F_2$ are given by \eqref{ext_rad_F(eta)}, $r_{\mathcal{C}}=r_{\mathcal{S}}$, $E_\mathcal{H}=\sqrt{1-2M/r_{\mathcal{H}}}$, $\eta_h=2 \arcsin E_\mathcal{H}$  and 
$\eta_{\mathcal{S}}=\arccos(2r_{\mathcal{S}}/r_{\mathcal{H}}-1)$. 
Let us find values for $r_{\mathcal{C}}$ and $r_{\mathcal{H}}$ such that
\begin{equation}
    \label{eq-ratio_coordinate_time_2}
    \Delta t_{\mathcal{C}_{\mathcal{S}\to\mathcal{A}}}(M,r_\mathcal{C})=  \Delta t_{\mathcal{U}_{\mathcal{S}\to\mathcal{A}}}(M,r_\mathcal{H},r_\mathcal{C})\,,
\end{equation}

The proper time  of the {observer} $\mathcal{U}$ is given by (cf. \eqref{ext_Deltatau_U})
\begin{equation}
\label{ext_Deltatau_U2}
    \Delta\tau_{\mathcal{U}_{\mathcal{S}\to\mathcal{A}}}(M, r_{\mathcal{S}}, r_{\mathcal{H}}) = \sqrt{\dfrac{r_{\mathcal{H}}^3}{2M}}(\eta_{\mathcal{S}} + \sin\eta_{\mathcal{S}})\,,
\end{equation}
where $r_{\mathcal{S}}=r_{\mathcal{C}}$, and $\Delta \tau_{\mathcal{C}}$ is given by \eqref{circ_deltatauc}, where $E_\mathcal{C}$ is given by \eqref{circ_E2}.

\begin{figure}[ht]
\centering
\includegraphics[width=10cm]{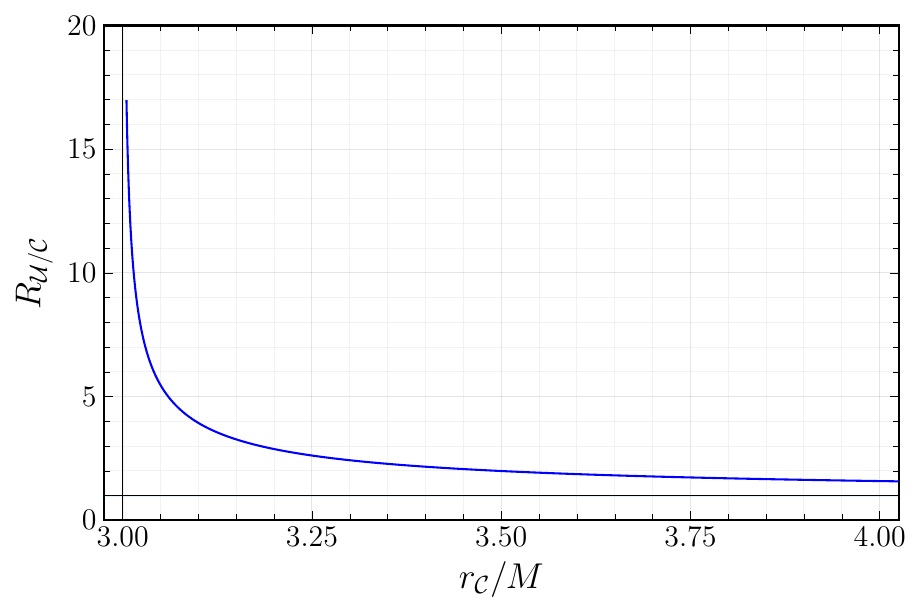}
\caption{
{Proper time ratio $R_{\mathcal{U}/\mathcal{C}}\coloneqq \Delta\tau_\mathcal{U}/ \Delta\tau_\mathcal{C}$ \emph{versus} the circular radius $r_\mathcal{C}/M = r_\mathcal{S}/M =r_\mathcal{A}/M$\,.}
}
\label{fig:ratio_U_over_C}
\end{figure}

The behavior of $R_{\mathcal{U}/\mathcal{C}}:=\Delta \tau_{\mathcal{U}}/\Delta \tau_{\mathcal{C}}$ in terms of $r_{\mathcal{C}}$  is shown in Fig.  \ref{fig:ratio_U_over_C}  In other words, observer $\mathcal{C}$ returns younger than $\mathcal{U}$, as expected. When $r_{\mathcal{C}}\rightarrow 3M$, $\Delta \tau_{\mathcal{U}}/\Delta \tau_{\mathcal{C}}\rightarrow \infty$.
\FloatBarrier

\subsection{Observers $\mathcal{D}$ and $\mathcal{C}$}
\label{subsec:D_and_C}

Finally, let us find the respective elapsed proper times of an \gls{observer} $\mathcal{C}$ in a circular orbit, at $r=r_{\mathcal{C}}>3M$, and an \gls{observer} $\mathcal{D}$ falling downward radially from $r=r_{\mathcal{C}}$, bouncing in a perfectly elastic way at $r=r_{\mathcal{B}}\in (2M,r_{\mathcal{C}})$ and returning to $r= r_{\mathcal{C}}$ to meet $\mathcal{C}$ (cf.~Fig.~\ref{fig:couples_of_twins}). To ensure that $\mathcal{C}$ and $\mathcal{D}$ meet again at the end of their trips, we need, analogously to what we did in the previous subsection, to adjust $r_{\mathcal{C}}$ and $r_{\mathcal{B}}$ accordingly so that the elapsed coordinate times for the round trips of $\mathcal{C}$ and $\mathcal{D}$  coincide. The coordinate time associated to one complete orbit  of $\mathcal{C}$ is given, according to  \eqref{circ_deltatc}, 
by
\begin{equation}
\label{circ_deltatc3}
\Delta t_{\mathcal{C}_{\mathcal{S}\to\mathcal{A}}}(M,r_\mathcal{C}) = 2\pi\sqrt{\frac{r_{\mathcal{C}}^3}{M}}\,,
\end{equation}
and the corresponding coordinate time for $\mathcal{D}$ to go down, bounce and return is, according to  \eqref{eq-coordinate_time_D}, given by
\begin{equation}
        \label{eq-coordinate_time_D2}
         \Delta t_{\mathcal{D}_{\mathcal{S}\to\mathcal{A}}}(M,r_{\mathcal{S}}, r_{\mathcal{B}})=\left[ 2E_\mathcal{S}\sqrt{\dfrac{r_{\mathcal{C}}^3}{2M}} F_1(\eta_\mathcal{B}, E_\mathcal{S}) + 4M F_2(\eta_\mathcal{B}, \eta_h) \right]\,,
        \end{equation}
where $F_1$ and $F_2$ are given by \eqref{ext_rad_F(eta)}, $E_\mathcal{S}=\sqrt{1-2M/r_{\mathcal{C}}}$, $\eta_h=2 \arcsin E_\mathcal{S}$  and 
$\eta_{\mathcal{B}}=\arccos(2r_{\mathcal{B}}/r_{\mathcal{C}}-1)$. 

\begin{figure}[ht]
\centering
\includegraphics[width=10cm]{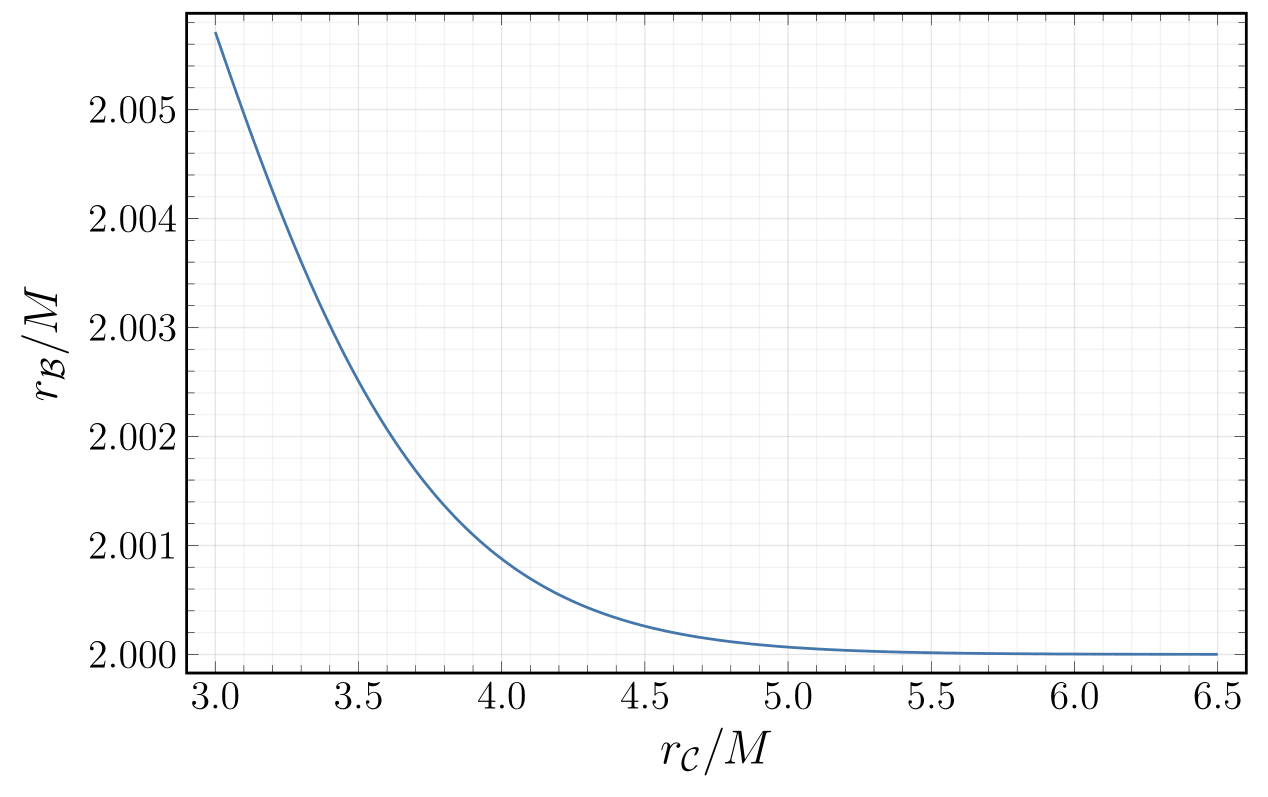}
\caption{{Bounce radius $r_\mathcal{B}{/M}$ of twin $\mathcal{D}$ \emph{versus} circular radius $r_\mathcal{C}{/M}$ of twin $\mathcal{C}$, under the constraint of equal elapsed coordinate times, along twins $\mathcal{D}$ and $\mathcal{C}$\,.}}
\label{fig:zeros_DC}
\end{figure}

\begin{figure}[ht]
\centering
\includegraphics[width=10cm]{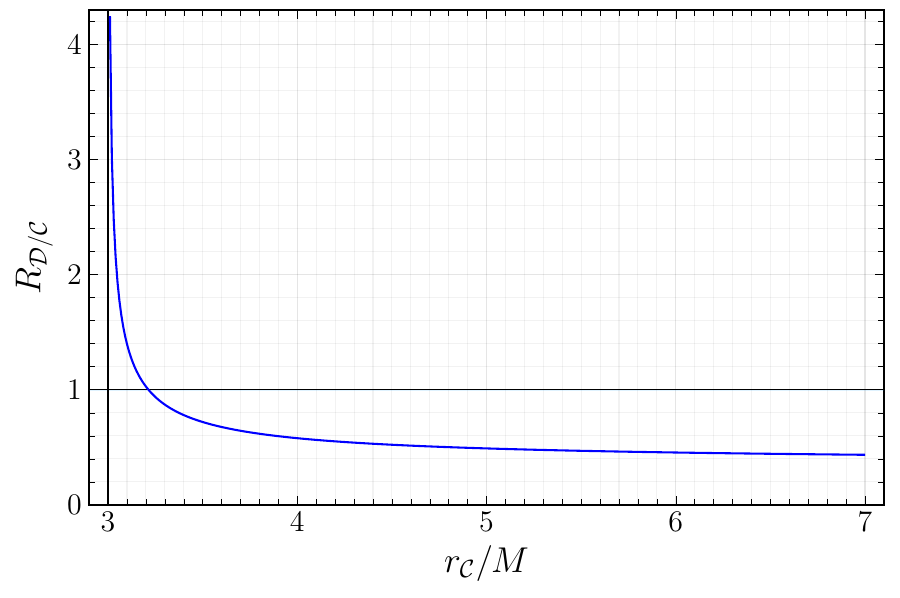}
\caption{{Proper time ratio  $R_{\mathcal{D}/\mathcal{C}}\coloneqq\Delta \tau_{\mathcal{D}} / \Delta \tau_{\mathcal{C}}$ \emph{versus} the circular radius $r_{\mathcal{C}}/M = r_\mathcal{S}/M = r_\mathcal{A}/M$\,.}}
\label{fig:ratio_D_over_C}
\end{figure}

Let us find values for $r_{\mathcal{B}}$ and $r_{\mathcal{C}}$ such that
\begin{equation}
    \label{eq-ratio_coordinate_time}
    \Delta t_{\mathcal{D}_{\mathcal{S}\to\mathcal{A}}}(M,r_{\mathcal{C}}, r_{\mathcal{B}})=  \Delta t_{\mathcal{C}_{\mathcal{S}\to\mathcal{A}}}(M,r_\mathcal{C})\,.
\end{equation}
Solving numerically the equation above shows that a pair of solutions $(r_{\mathcal{C}},r_{\mathcal{B}})$ always exists, albeit for $r_{\mathcal{B}}$ very close to the horizon $r_h=2M$. As $r_{\mathcal{C}}$ grows, $r_{\mathcal{B}}\rightarrow 2M$, as shown in Fig. \ref{fig:zeros_DC}.
The proper times of the {observers} $\mathcal{C}$ and  $\mathcal{D}$ are given by \eqref{circ_deltatauc}  and \eqref{eq-proper-time-down} (with $r_{\mathcal{S}}=r_{\mathcal{C}}$). In Fig.  \ref{fig:ratio_D_over_C}, we observe the surprising result that {observer} $\mathcal{D}$ may return younger or older than $\mathcal{C}$, depending on $r_{\mathcal{C}}/M$. For $r_{\mathcal{C}}/M\approx 3.21337$ both observers are at the same age at the end of the journey. This reversal of differential aging is, however, to be expected since, as $r_{\mathcal{C}}\rightarrow 3M$, the circular geodesic $\mathcal{C}$ becomes arbitrarily close to a null geodesic, causing its elapsed proper time to approach zero.

In contrast, it is worth mentioning that if we try, in the Newtonian case, to arrange the observers $\mathcal{C}$ and $\mathcal{D}$ to reunite after an exact orbital period of $\mathcal{C}$ and a bounce of $\mathcal{D}$, this turns out to be impossible.

This example shows that the fountain of youth argument holds only if the \gls{observer} compared with $\mathcal{Y}$ is the static \gls{observer} $\mathcal{K}$; for a general observer outside the fountain of youth, the observer going inside it may return older. One could argue that $\mathcal{C}$ is also ``inside'' the fountain of youth region, as it is exactly in its exterior bounding surface. However, as $\mathcal{C}$ is at the boundary, we may add measure-zero radial curves that throw it away from the fountain of youth without affecting the total proper time elapsed.
\FloatBarrier

\section{Universal differential aging and Doppler effect?}
\label{sec:Doppler_effect}

A connection between differential aging and the Doppler effect is often observed in the literature \cite{Rindler2006b, Schutz2022}. We want to take the opportunity here to make some remarks on this issue, which is also connected to the observation regarding the measuring of times by \gls{observer} in Sec. \ref{sec:differential_aging}. In fact, it will turn out to be expedient to distinguish between two kinds of Doppler effect in a generic curved spacetime.

\begin{figure}
	\centering\includegraphics[width=12cm]{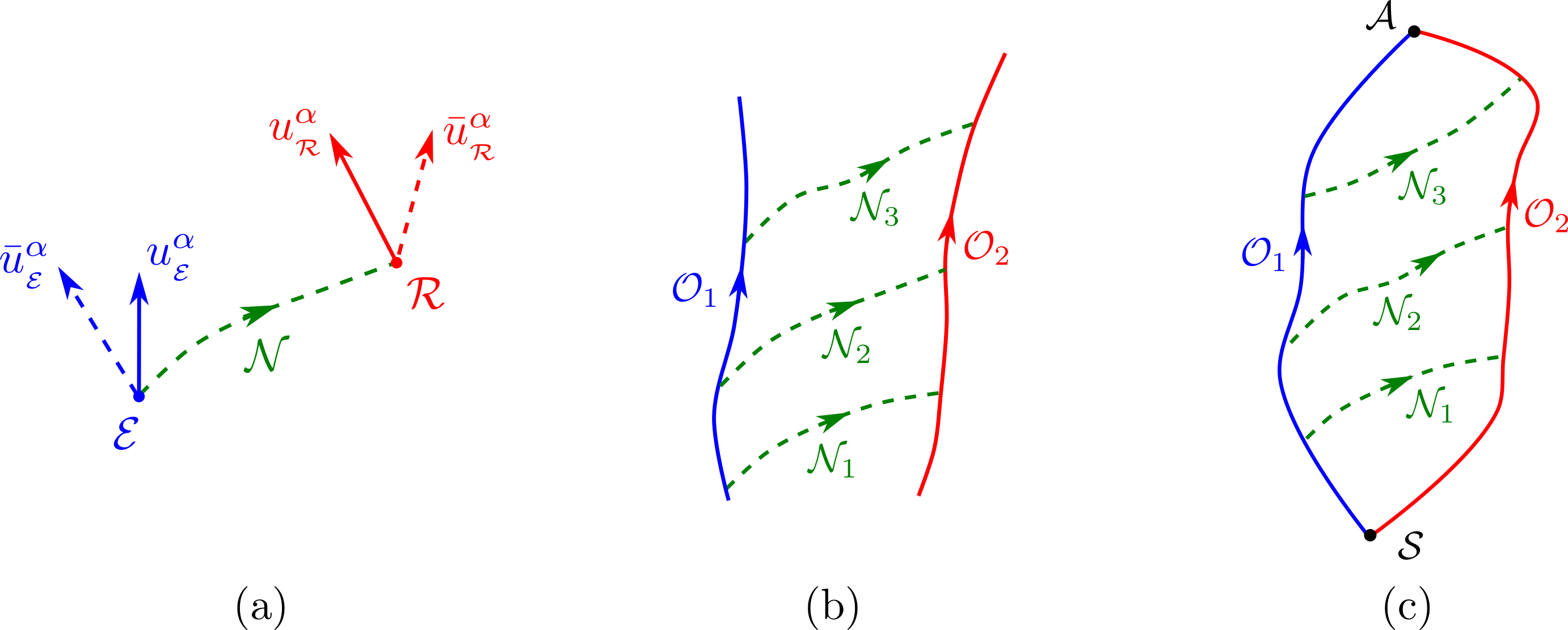}
	\caption{(a) Doppler effect between \glspl{instantaneous_observer}; (b) Doppler effect between \glspl{observer}, and (c) relationship between Doppler effect and differential aging.}
	\label{fig:all_that}
\end{figure}

The \emph{Doppler effect between two instantaneous observers} [cf. Fig. \ref{fig:all_that}, panel (a)] involves the comparison of the wavelengths measured by exactly two \glspl{instantaneous_observer}, $(\mathcal{E}, u^\alpha_\mathcal{E})$ and $(\mathcal{R}, u^\alpha_\mathcal{R})$\,, defined at an emission event $\mathcal{E}$ and a reception event $\mathcal{R}$\,, connected by a null geodesic $\mathcal{N}$\,, in an arbitrary \gls{spacetime} $\left(\mathcal{M}, g_{\alpha\beta}\right)$\,. The Doppler shift is defined by
\begin{align}
    z[u^\alpha_\mathcal{E}\to u^\alpha_\mathcal{R}] &\coloneqq \dfrac{\Delta\lambda}{\lambda_\mathcal{E}} \coloneqq \dfrac{\lambda_\mathcal{R} - \lambda_\mathcal{E}}{\lambda_\mathcal{E}}\,, 
\end{align}
or, equivalently,
\begin{equation}
    1 + z[u^\alpha_\mathcal{E}\to u^\alpha_\mathcal{R}] = \dfrac{\lambda_\mathcal{R}}{\lambda_\mathcal{E}} = \dfrac{\left.\left(k^\alpha u_\alpha\right)\right|_\mathcal{R}}{\left.\left(k^\beta u_\beta\right)\right|_\mathcal{E}}\,. \label{fundamental_redshift}
\end{equation}
If we change the \glspl{instantaneous_observer}: $(\mathcal{E}, u^\alpha_\mathcal{E})\mapsto (\mathcal{E}, \bar{u}^\alpha_\mathcal{E})$\,, $(\mathcal{R}, u^\alpha_\mathcal{R})\mapsto (\mathcal{R}, \bar{u}^\alpha_\mathcal{R})$\,, the Doppler shift changes as
\begin{align}
    1 + z[\bar{u}^\alpha_\mathcal{E}\to \bar{u}^\alpha_\mathcal{R}] &= \left(1 + z[\bar{u}^\alpha_\mathcal{E}\to u^\alpha_\mathcal{E}]\right)\left(1 + z[u^\alpha_\mathcal{E}\to u^\alpha_\mathcal{R}]\right)\left(1 + z[u^\alpha_\mathcal{R}\to \bar{u}^\alpha_\mathcal{R}]\right)\,,
\intertext{or in a lighter (but not so explicit) notation,}
   1 + \bar{z}_{\mathcal{E}\to\mathcal{R}} &= \left(1+z_{\text{loc},\mathcal{E}}\right)\left( 1 + z_{\mathcal{E}\to\mathcal{R}}\right)\left(  
 1 + z_{\text{loc},\mathcal{R}}\right)\,. \label{redshift_transformation}
\end{align}

\indent Some observations are in order: (i) in a general \gls{spacetime} the global Doppler shifts ($\bar{z}_{\mathcal{E}\to\mathcal{R}}$ and $z_{\mathcal{E}\to\mathcal{R}}$) cannot be unambiguously conceived as arising from two objective, absolute contributions: one kinematic and another gravitational (or cosmological). Only when there is a class of privileged \glspl{instantaneous_observer} (arising, e.g., from some \gls{isometry}, such as in Schwarzschild, Kerr or Robertson-Walker spacetimes) does such a decomposition make sense; (ii) in this pure context, one cannot relate the Doppler effect to any putative differential aging at all, since we do not even have full-fledged \glspl{observer} to begin with.

Panel (b) of Fig. \ref{fig:all_that}, in contrast, illustrates the \emph{Doppler effect between two observers}, $\mathcal{O}_1$ and $\mathcal{O}_2$. Here we are again unable to directly relate this situation to the differential aging since, although there are now indeed two \glspl{observer}, they do not meet.

Only in the situation illustrated in Fig.~\ref{fig:all_that}, panel (c), where the \glspl{observer} now do meet, does it make sense to speak about both the redshift between instantaneous observers along $\mathcal{O}_1$ and $\mathcal{O}_2$ and their differential aging. Granting that, we might then inquire whether there is a universal relationship, in a generic \gls{spacetime}, between these phenomena. 

\begin{figure}[ht]
	\centering
	\includegraphics[width=5.5cm]{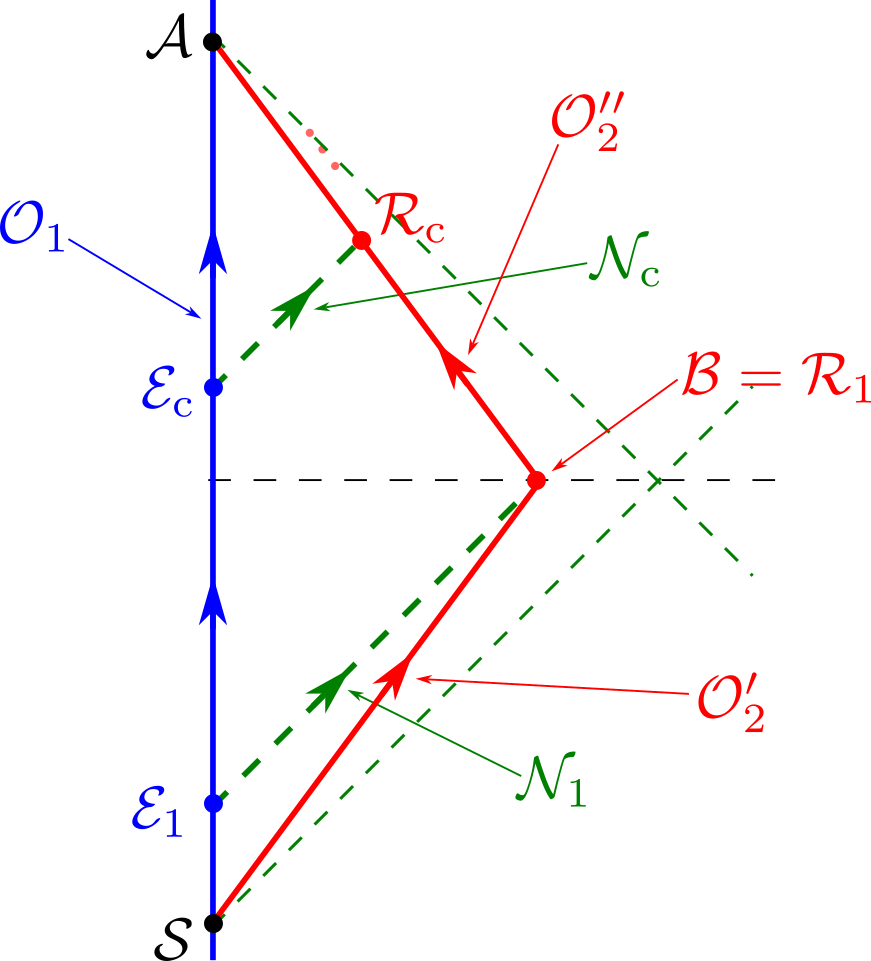}
	\caption{Prototypical differential aging scenario in Minkowski spacetime. $\mathcal{O}_1$ is a differentiable geodesic \gls{observer} and  $\mathcal{O}_2=\mathcal{O}'_2\cup\mathcal{O}''_2$ is a broken (piecewise differentiable) geodesic \gls{observer}, constituted by two geodesic {observers} $\mathcal{O}'_2$ and $\mathcal{O}''_2$\,, with a common bounce (or break) event $\mathcal{B}$\,, which have the same (3-)speed relative to $\mathcal{O}_1$, but in opposite directions. Thus there is a single null geodesic $\mathcal{N}_1$\,, emitted by $\mathcal{O}_1$\,, at event $\mathcal{E}_1$\,, which connects to the reception event $\mathcal{R}_1\coloneqq\mathcal{B}$\,, at $\mathcal{O}_2$\,. }
	\label{fig:prototypical_twin_paradox_in_Minkowski_spacetime}
\end{figure}

For the prototypical case of the differential aging in Minkowski \gls{spacetime} shown in Fig.  \ref{fig:prototypical_twin_paradox_in_Minkowski_spacetime}, it is apparent, from the monitoring of the Doppler shift between the inertial \gls{observer} $\mathcal{O}_1$ and the broken inertial \gls{observer} $\mathcal{O}_2 = \mathcal{O}'_2\cup\mathcal{O}''_2$\,, that $\mathcal{O}_2$ gets younger than $\mathcal{O}_1$\,.

\FloatBarrier

The proper time $\tau_{{\mathcal{O}_1}_{{\mathcal{S}}\to\mathcal{E}_1}}$ elapsed along $\mathcal{O}_1$\,, from $\mathcal{S}$ to $\mathcal{E}_1$, is shorter than the proper time $\tau_{{\mathcal{O}_2}_{{\mathcal{S}}\to\mathcal{R}_1}}$ elapsed along $\mathcal{O}_2$\,, from $\mathcal{S}$ to $\mathcal{R}_1$\,, because of their \emph{redshift}. Now, by reciprocity, in the approaching stage, there is the same ratio of proper times, but now giving rise to a \emph{blueshift}. Thus the proper time $\tau_{{\mathcal{O}_1}_{\mathcal{E}_1\to\mathcal{A}}}$ elapsed along $\mathcal{O}_1$\,, from $\mathcal{E}_1$ to $\mathcal{A}$ is longer than the proper time $\tau_{{\mathcal{O}_2}_{{\mathcal{R}_1}\to\mathcal{A}}}$ elapsed along $\mathcal{O}_2$\,, from $\mathcal{R}_1$ to $\mathcal{A}$\,, by the very same ratio. Moreover, by the symmetry of this prototypical problem, we have
\begin{equation}
\tau_{{\mathcal{O}_2}_{{\mathcal{S}}\to\mathcal{R}_1}}=\tau_{{\mathcal{O}_2}_{{\mathcal{R}_1}\to\mathcal{A}}}=x
\end{equation}
On the other hand, we may write:
\begin{equation}
\tau_{{\mathcal{O}_1}_{{\mathcal{S}}\to\mathcal{E}_1}} = K\tau_{{\mathcal{O}_2}_{{\mathcal{S}}\to\mathcal{R}_1}} = Kx\quad;\quad\tau_{{\mathcal{O}_1}_{\mathcal{E}_1\to\mathcal{A}}}=\frac{1}{K}\tau_{{\mathcal{O}_2}_{{\mathcal{R}_1}\to\mathcal{A}}} = \frac{x}{K}\;,
\end{equation}
where $0 < K < 1$. Finally, we get:
\begin{equation}
    \tau_{{\mathcal{O}_1}_{{\mathcal{S}}\to\mathcal{A}}} = \tau_{{\mathcal{O}_1}_{{\mathcal{S}}\to\mathcal{E}_1}} + \tau_{{\mathcal{O}_1}_{\mathcal{E}_1\to\mathcal{A}}} = \left(K + \frac{1}{K}\right)x > 2x = \tau_{{\mathcal{O}_2}_{{\mathcal{S}}\to\mathcal{A}}}\;.
\end{equation}
Thus, we see that, directly based on a Doppler shift argument, the twin $\mathcal{O}_1$ indeed gets older than the twin $\mathcal{O}_2$. Of course, this result is much more easily obtained via the expression for the interval in pseudo-Cartesian coordinates adapted to the inertial {observer} $\mathcal{O}_1$\,.

However, in a generic spacetime, and even in Schwarzschild spacetime, things are much subtler, as we have remarked in the counterexamples of Sec. \ref{sec:comparison_proper_times}. In fact, using Doppler shift formulas for \glspl{observer} $\mathcal{K}$, and $\mathcal{U}$, it can be shown  (e.g., \cite{Kasai2023}) that we can have blueshift and redshift at different stages of the journey. Intuitively, there will always be an initial stage in which the twins measure a \emph{redshift} and then eventually a final stage in which they measure a \emph{blueshift}, for after all they are initially ``receding'' and then eventually ``approaching''. However, in an arbitrary (curved) \gls{spacetime} context, in contrast to that prototypical Minkowski spacetime case, we no longer have such a high symmetry of the {observers} and a vanishing curvature. Of course, from Eqs. \eqref{fundamental_redshift} and \eqref{redshift_transformation}, we can always choose {instantaneous observers}, at emission and reception events, which are parallel transported (along the connecting null geodesic) and this implies the corresponding global Doppler shift $z_{\mathcal{E}\to\mathcal{R}}$ vanishes; one then only needs to take into account the local (pointwise) Doppler shifts at emission, $z_{\text{loc},\mathcal{E}}$\,,  and reception, $z_{\text{loc},\mathcal{R}}$\,, via usual Lorentz boosts [cf. Fig.~\ref{fig:all_that}, panel (a)]. This does not provide much insight because the final global Doppler shift $\bar{z}_{\mathcal{E}\to\mathcal{R}}$ is determined by the aforementioned boosts, which camouflage the curvature/metric via the parallel transportation.

We claim the bottom line is that, in a generic, arbitrary spacetime, to compare the elapsed proper times along two arbitrary {observers}, there is no easy replacement for the calculation of the corresponding line integrals; in some particular cases, one may resort to local theorems (e.g., \cite{Sokolowski2017}).

\section{Discussion}
\label{sec:discussion}

We have systematically studied the differential aging in a vacuum Schwarzschild spacetime, deriving the proper time intervals for four convenient \glspl{observer} (``twins''): a static Killing one ($\mathcal{S}$), two smooth geodesic ones ($\mathcal{U}$ and $\mathcal{C}$) and a sectionally geodesic one ($\mathcal{D}$). We plotted all six independent corresponding proper time ratios, and we now present a summary of the qualitative results in Table \ref{tab:couples_and_differential_agings}. 

\begin{table}
	\centering
	\begin{tabular}{|c|c|}
		\hline
                           & \\
		Twins $[(\mathcal{O}_1, \mathcal{O}_2)]$ & Proper time ratio $R_{\mathcal{O}_1/\mathcal{O}_2}$ [Fig.] \\
                   & \\
		\hline
              & \\
                $(\mathcal{U}, \mathcal{K})$ & $\geq 1$\qquad [\ref{fig:ratio_U_over_K}] \\
               & \\
                $(\mathcal{K}, \mathcal{D})$ & $\geq 1$\qquad [\ref{fig:ratio_D_over_K}] \\
               & \\
                $(\mathcal{U}, \mathcal{D})$ & $\geq 1$\qquad [\ref{fig:ratio_D_over_U}] \\
                & \\
                $(\mathcal{U}, \mathcal{C})$ & $\geq 1$\qquad [\ref{fig:ratio_U_over_C}] \\
               & \\
                $(\mathcal{C}, \mathcal{K})$ & $\leq 1$\qquad [\ref{fig:ratio_C_over_K}] \\
                & \\               
                $(\mathcal{C}, \mathcal{D})$ & $\lesseqgtr 1$\qquad [\ref{fig:ratio_D_over_C}] \\
                & \\
		\hline
	\end{tabular}
	\caption{{Twins $(\mathcal{O}_1, \mathcal{O}_2)$ and corresponding proper time ratio $R_{\mathcal{O}_1/\mathcal{O}_2}\coloneqq\Delta\tau_{\mathcal{O}_1}/\Delta\tau_{\mathcal{O}_2}$ (cf. the cited Figure for a relevant plot of $R_{\mathcal{O}_1/\mathcal{O}_2}\times r_*/M$\,, where $r_*$ is a relevant radius which characterizes the arrangement of the twins).}}
	\label{tab:couples_and_differential_agings}
\end{table} 

The key takeaway points we want to stress are conveniently summarized in Table \ref{tab:couples_and_differential_agings}, which suggests, by means of examples in its first four rows, for the twins $(\mathcal{U}, \mathcal{K})$\,, $(\mathcal{K}, \mathcal{D})$\, $(\mathcal{U}, \mathcal{D})$\, and $(\mathcal{U}, \mathcal{C})$\,, that the twin that travels closer to the matter source indeed gets younger than the one which stays further away. However this same table, in its last two rows, provides two enlightening and silver-bullet counterexamples to two diehard myths: (i) for the twins $(\mathcal{D}, \mathcal{C})$\, there is a range of initial conditions such that the twin $\mathcal{D}$, which is always closer to the source, returns \emph{older} than the twin $\mathcal{C}$\,, which always remains further away; (ii) the geodesic twin $\mathcal{C}$ gets \emph{younger} than the accelerated twin $\mathcal{K}$, despite being always at the same radius, and thus somehow subjected to the ``same'' gravitational field.  It is in fact true that the counterexample we found here for {observer} $\mathcal{C}$ is not very realistic, since its radius has to be {close to $3M$, i.e., inside the interval $3M<r<6M$, being therefore an unstable orbit.}
Moreover, if we choose, at a certain event of the curve $\mathcal{C}$, the associated \gls{instantaneous_observer} $\bar{u}^\alpha$, given by [cf. \eqref{circ_deltatc} and \eqref{circ_deltatauc}]
\begin{equation}\label{circinstobs}
    \bar{u}^\alpha = \left. \dfrac{dx^\alpha}{d\tau}\right|_{\mathcal{C}}\,,
\end{equation}
and take the Killing {instantaneous observer} $u^\alpha$ defined, at the same event, by
\begin{equation}\label{killinginstobs}
    u^\alpha = \dfrac{1}{\sqrt{1-2M/r}}\,\delta^\alpha_0\,,
\end{equation}
we can calculate the relative speed $v$ between them via
\begin{equation}\label{gammalorentz}
    \gamma(v):= \dfrac{1}{\sqrt{1 - v^2}} = -\bar{u}^\alpha {u}_\alpha\,.
\end{equation}
 Indeed, this speed is very close to the speed of light, as we now show. From \eqref{circinstobs}  to \eqref{gammalorentz}  :
\begin{equation}
    \gamma(v) = \sqrt{-g_{00}}\,\left.\frac{dx^0}{d\tau}\right|_\mathcal{C}=\sqrt{1-\frac{2M}{r_\mathcal{C}}}\left.\frac{dt}{d\tau}\right|_\mathcal{C}\;.
\end{equation}
Using  \eqref{circ_deltatauc} and \eqref{circ_E2}, we also have:
\begin{equation}
    \left.\frac{dt}{d\tau}\right|_\mathcal{C} = \frac{E_\mathcal{C}}{1-2M/r_\mathcal{C}} = \frac{1}{\sqrt{1-3M/r_\mathcal{C}}}\;.
\end{equation}
Comparing the above expressions:
\begin{equation}
    \frac{1}{\sqrt{1-v^2}} = \frac{\sqrt{1-2M/r_\mathcal{C}}}{\sqrt{1-3M/r_\mathcal{C}}} \quad\Longrightarrow\quad v = \sqrt{\frac{M}{r_\mathcal{C}-2M}}\;.
\end{equation}
Setting the radius $r_\mathcal{C} \approx 3.21337M$, where both {observers} have the same age (cf. \ref{subsec:D_and_C}), the corresponding relative speed turns out to be $v\approx0.90783$. Therefore, for the observer $\mathcal{D}$ to return older than the observer $\mathcal{C}$, $v > 0.90783$. This is expected since $r=3M$ corresponds to the innermost unstable circular null geodesic.

The deceptive claim that time runs slower when we are closer to a mass source is also recurrent in the problem of the Doppler effect. We saw in Sec. \ref{sec:Doppler_effect} that the usual problem of the Doppler effect is, in principle, different from the issue of the differential aging. We may draw a correspondence between them only when the \glspl{observer} of the former meet in two different events. 
 
As we have seen, even in this situation, the conclusions concerning the aging of observers are not immediate --- in the problem described at the end of Sec. \ref{sec:Doppler_effect}, we made strong use of symmetries both of the Minkowski metric and of the observers, which may be absent in more general situations. In fact, even in the still highly symmetric Schwarzschild case, for instance, among \glspl{observer} $\mathcal{U}$ and $\mathcal{K}$, the relation between the Doppler shift and differential aging is not clear.

A natural follow-up to this work would be extending this analysis to other observers still in the vacuum Schwarzschild spacetime \cite{Fung2016}, or to the region inside the vacuum Schwarzschild event horizon, or to other spacetimes (such as has already been done for, e.g., Kerr \cite{Markley1973} and de Sitter spacetimes \cite{Boblest2011}).

We have created a GitHub repository\footnote{\url{https://github.com/fernandodeeke/Twin\_Paradox}}, where the reader may find interactive Python notebooks to illustrate the ratio of proper times for the distinct pairs of observers.

\printglossaries

\bibliographystyle{unsrt}
\bibliography{twin_paradox}

\end{document}